\documentclass[12pt]{article}
\usepackage{array}
\usepackage{graphicx}
\usepackage{amssymb}
\usepackage{amsmath}
\usepackage{multirow}
\input{epsf}
\usepackage{cite}
\def\@fmsl@sh#1#2#3{\m@th\ooalign{$\hfil#1\mkern#2/\hfil$\crcr$#1#3$}}
 \def\eq#1\en{\begin{equation}#1\end{equation}}
\def\s[#1,#2]{[#1\stackrel{\star}{,}#2]}
\def\sx[#1,#2]{[#1\stackrel{\star_{x}}{,}#2]}

\newcommand{\nc}{\newcommand}
\nc{\beq}{\begin{equation}}
\nc{\eeq}{\end{equation}}
\nc{\beqa}{\begin{eqnarray}}
\nc{\eeqa}{\end{eqnarray}}

\def\bc{\begin{center}}
\def\ec{\end{center}}

\def\to{\rightarrow}

\def\gsim{\mathrel{\mathpalette\atversim>}}

\def\bc{\begin{center}}
\def\ec{\end{center}}

\def\gsim{\mathrel{\rlap{\lower4pt\hbox{\hskip1pt$\sim$}}

    \raise1pt\hbox{$>$}}}       

\def\gsim{\mathrel{\rlap{\lower4pt\hbox{\hskip1pt$\sim$}}
    \raise1pt\hbox{$>$}}}       

\begin{document}
 
\makeatletter
\def\fmslash{\@ifnextchar[{\fmsl@sh}{\fmsl@sh[0mu]}}
\def\fmsl@sh[#1]#2{%
  \mathchoice
    {\@fmsl@sh\displaystyle{#1}{#2}}%
    {\@fmsl@sh\textstyle{#1}{#2}}%
    {\@fmsl@sh\scriptstyle{#1}{#2}}%
    {\@fmsl@sh\scriptscriptstyle{#1}{#2}}}
\def\@fmsl@sh#1#2#3{\m@th\ooalign{$\hfil#1\mkern#2/\hfil$\crcr$#1#3$}}
\makeatother

\thispagestyle{empty}
\begin{titlepage}
\boldmath
\begin{center}
  \Large {\bf Non-thermal quantum black holes with quantized masses}
    \end{center}
\unboldmath
\vspace{0.2cm}
\begin{center}
{ {\large Xavier Calmet}\footnote{x.calmet@sussex.ac.uk}
 and
  {\large Nina Gausmann}\footnote{n.gausmann@sussex.ac.uk}
}
 \end{center}
\begin{center}
{\sl Physics $\&$ Astronomy, 
University of Sussex, Falmer, Brighton, BN1 9QH, UK 
}
\end{center}
\vspace{\fill}
\begin{abstract}
\noindent
In this paper we discuss non-thermal quantum black holes with a discrete mass spectrum and their possible new signatures at the LHC. We calculate the inclusive cross sections for the production of quantum black holes with discrete masses at the LHC as well as some exclusive cross sections for particularly interesting decay modes. 
\end{abstract}

{\it keywords:} low scale quantum gravity, quantum black holes, LHC
\end{titlepage}


\subsubsection*{Introduction}

The advent of models with low scale quantum gravity designed to address the hierarchy problem  \cite{ArkaniHamed:1998rs,Randall:1999ee,Calmet:2008tn}, see e.g. \cite{Calmet:2010nt} for a recent review,  has led to much progress in the physics of small black holes which could be produced in high energy collisions of  particles.  Black holes created in such processes are very different from their astrophysical counterparts, indeed their masses are expected to be close to the Planck scale which in the models aforementioned is in the TeV region. Eardley and Giddings \cite{Eardley:2002re}, following the works of Payne and D'Eath \cite{D'Eath:1992hb} and Penrose (unpublished) have established that even for a non zero impact parameter, a classical black hole would form if one collided two particles in the limit of a very large center of mass energy. A semi-classical formulation of this construction using a path integral formalism appeared in  \cite{Hsu:2002bd}. Hsu demonstrated that for collisions with center of 
mass energies 5 to 20 times 	larger than the Planck scale, depending on the model \cite{Meade:2007sz}, a semi-classical black hole will form. Semi-classical black holes are thermal objects, their decays are expected to be well described by Hawking radiation.

The production of microscopic black holes at colliders such as the Large Hadron Collider (LHC) would provide one of the most interesting features of low scale quantum gravitational effects and has now been extensively studied for semi-classical black holes \cite{Dimopoulos:2001hw,Banks:1999gd,Giddings:2001bu,Feng:2001ib,Anchordoqui:2003ug,Anchordoqui:2001cg,Anchordoqui:2003jr,Dai:2007ki}.  However,  even in the most optimistic case of a few TeV Planck mass, it is clear that the number of semi-classical black holes that could be produced at the LHC is very limited since the first semi-classical black hole would have a mass between 5 TeV and 20 TeV for a 1 TeV Planck scale.

On the other hand, it was pointed out in \cite{Calmet:2008dg} that non-thermal quantum black holes could be of particular importance. Because their masses are so close to the Planck scale, their decomposition is not expected to be described accurately by Hawking radiation. One expects that these black holes will rather decay into a couple of particles. These black holes can be classified according to the quantum numbers of the particles that formed them \cite{Calmet:2008dg}. It is thus easy to predict their decomposition modes and to calculate the corresponding branching ratios. These events may resemble strong gravitational rescattering processes  \cite{Meade:2007sz}.


\subsubsection*{Production cross sections of QBHs with continuous and discrete mass spectra}

Till recently, only a continuous mass spectrum had been considered for quantum black holes. This was sensible since most of the physics is extrapolated from the classical black hole case. However, motivated by the existence of a minimal length in models trying to unify quantum mechanics with general relativity \cite{min}, it was suggested in \cite{Calmet:2012cn} (see also \cite{Dvali:2011nh}) that non-thermal quantum black hole masses could be quantized in terms of the Planck scale. The phenomenology of these black holes at the LHC would be rather different from those with a continuous mass spectrum since their mass spectrum is discrete.\\
In the case of a continuous mass spectrum, the cross section for non-thermal black holes \cite{Calmet:2008dg} was extrapolated from the semi-classical black hole cross section \cite{Anchordoqui:2003jr}
  \begin{eqnarray}
    \sigma^{pp}(s,x_{min},n,M_D) &=& \int_0^1 2z dz \int_{\frac{(x_{min} M_D)^2}{y(z)^2 s}}^1 du \int_u^1 \frac{dv}{v}  \\ \nonumber && \times F(n) \pi r_s^2(us,n,M_D) \sum_{i,j} f_i(v,Q) f_j(u/v,Q)
  \end{eqnarray}
where $\sqrt{s}$ is the centre of mass (CoM) energy, $n$ is the number of extra dimensions, $z \equiv b/b_{max}$ is the rescaled impact parameter,
$F\left( n \right)$ and $y(z)$ describe the effects of inelasticity numerically fitted by Yoshino and Nambu \cite{Yoshino:2002tx,Yoshino2}. The labels
$i$,$j$ run over the different particles species, $f_{i}$,$f_{j}$ are parton distribution functions evaluated at the scale of momentum transfer $Q$, $u$ and $v$ are the momentum fractions of the incoming particles,
and $x_{min}=M_{BH}^{min}/\text{M}_{\text{D}}$ is the ratio of the threshold of the black hole production $M_{BH}^{min}$ over the reduced Planck mass. 
The $n$ dimensional  Schwarzschild radius is given by
  \begin{eqnarray}
    r_s(us,n,M_D)=k(n)M_D^{-1}[\sqrt{us}/M_D]^{1/(1+n)}
  \end{eqnarray}
with 
  \begin{eqnarray}
  k(n) =  \left [2^n \sqrt{\pi}^{n-3} \frac{\Gamma((3+n)/2)}{2+n} \right ]^{1/(1+n)} \,.
  \end{eqnarray}
  On the other hand, if the mass spectrum is discrete, the cross section is given by  \cite{Calmet:2012cn}: 
  \begin{eqnarray}
  \label{pp_total}
  \sigma^{pp}_{tot}(s,n,M_D) &=& \sum_i \sigma^{pp}_{QBH}(s,M^i_{QBH},n,M_D) \,. 
  \end{eqnarray}
In the case of a discrete mass spectrum, one could thus describe QBHs as heavy particle states with a very short life time ($\Gamma (QBH\to 2\  {\rm particles})\sim \frac{1}{64 \pi^2} M_{QBH}$ using dimensional analysis). Each individual production cross section in the discrete case has the following form:
  \begin{eqnarray}
  \label{pp_individual}
  \sigma^{pp}_{QBH}(s,M_{QBH},n,M_D) &=&  \pi r_s^2(M_{QBH}^2,n,M_D) \int_0^1 2z dz \int_{\frac{(M_{QBH})^2}{y(z)^2 s}}^1 du \int_u^1 \frac{dv}{v}  \\ \nonumber && \times F(n)  \sum_{i,j} f_i(v,Q) f_j(u/v,Q)
  \end{eqnarray}
where $M_{QBH}$ is the mass of the black hole, respectively. For the Schwarzschild radius, we take
  \begin{eqnarray}
  r_s(M^2_{QBH},n,M_D)=k(n)M_D^{-1}[M_{QBH}/M_D]^{1/(1+n)}
  \end{eqnarray}
and $k(n)$ remains the same as in the continuous case. Assuming that space-time is quantized at short distances and that there is  a minimal length  \cite{min,mead,Padmanabhan:au,Calmet:2004mp,Calmet:2005mh,Calmet:2007vb}, the mass distribution is expected to be partitioned in terms of the Planck mass. If the Planck mass is of the order of 1 TeV, one would expect five different black hole states before reaching a continuum in the semi-classical regime. The factors $F(n)$ and the functions $y(z)$ have been calculated in the framework of classical black holes and are not known for non-thermal quantum black holes. We are thus setting the factors $F(n)=1$ as well as the functions $y(z)=1$. This is an approximation which is reasonable since these factors have been calculated for classical black holes and may not apply to quantum black holes. However, let us emphasize that the functions $y(z)$ describe the lost energy of the incoming partons to gravitational radiation. Since this has nothing to do with the 
decomposition of the black holes or the type of black hole produced, we expect that the functions $y(z)$ for quantum black holes will be similar to the classical case.

One might worry that the cross section for  non-thermal black holes could be exponentially suppressed. However, such a suppression has been refuted in \cite{Hsu:2002bd}: Because of the seminal work of Eardley and Giddings \cite{Eardley:2002re}, we know that  there are classical trajectories with two particle initial conditions which evolve into black holes, the process is clearly not classically forbidden and hence there is no tunneling factor. Furthermore, the time-reversed process is not thermal and involves very special initial (final) conditions. 


\subsubsection*{Classification of QBHs}

To classify a quantum black hole, we need to consider not only its mass and spin but also its charge under  $\text{U}(1)_{\text{em}}$ and $\text{SU}(3)_{\text{c}}$. The fact that quantum black holes carry a color charge is not in contradiction with the notion of confinement because the production and decay of these black holes happen over a short space-time length which is much shorter than the hadronization length scale. In the case of the LHC,  protons are collided with  protons, thus the quantum black hole will only be created by collisions of quarks, anti-quarks, and gluons. This yields the following possible color states for the quantum black holes, depending on the initial particles: 
  \begin{itemize}
    \item[a)] $\text{q} + \bar{\text{q}}: \ 			{\bf 3} \times {\bf \overline 3}= {\bf 8} + {\bf 1}$
    \item[b)] $\text{q} + \text{q}: \ 					{\bf 3} \times {\bf 3}= {\bf 6} + {\bf \overline 3}$
    \item[c)] $\bar{\text{q}} + \bar{\text{q}}: \	{\bf \overline 3} \times {\bf \overline 3}= {\bf \overline 6} + {\bf 3}$
    \item[d)] $\text{q} + \text{g}: \ {\bf 3} \times {\bf 8}= {\bf 3} + {\bf \overline 6}+ {\bf 15}$
    \item[e)] $\bar{\text{q}} + \text{g}: \ {\bf \overline 3} \times {\bf 8}= {\bf \overline 3} + {\bf 6}+ {\bf \overline{15}}$
    \item[f)] $\text{g} + \text{g}: \ {\bf 8} \times {\bf 8}= {\bf 1}_S + {\bf 8}_S+ {\bf 8}_A+{\bf 10} + {\bf \overline{10}}_A+ {\bf 27}_S$ \, .
  \end{itemize}
There are nine possible electromagnetic charges ($0$, $\pm 1/3$,  $\pm 2/3$, $\pm 4/3$, $\pm 1$), corresponding to the charges of the colliding particles. Due to the high center-of-mass energies, we consider all initial particles to be massless. Using Clebsch-Gordon-coefficients\cite{pdg}, we find that in a collision of two fundamental fermions ($1/2 \times 1/2$), the spin-1 state is three times more likely to form than the spin-0 state. If a massless vectorboson collides with a fundamental fermion, the spin-$1/2$ state is just half as likely to form compared to the spin-$3/2$ state and for two gluons the spin state ratio for the spin-0, spin-1 and spin-2 is 2:3:7. All spin factors are displayed in Table (\ref{spin}).
  \begin{table}[!ht] 
    \begin{center}
      \caption{\textbf{Spin factors} for massless particles, determined by Clebsch-Gordon-coefficients  \label{spin}}
      \begin{tabular}[c]{lc|c|c|}
        &  \multicolumn{1}{c}{$0$}  & \multicolumn{1}{c}{$1/2$} &  \multicolumn{1}{c}{$1$}      \\
        \cline{2-4}
        \multicolumn{1}{r|}{$0$}                     &  $0$  & $1/2$     &  $1$                 \\
        \cline{2-4}
        \multicolumn{2}{r|}{\multirow{2}{*}{$1/2$}}          & $0\,,\,1$ &  $1/2\,,\,3/2$       \\
        \multicolumn{2}{r|}{ }                               & $1:3$     &  $1:2$               \\
        \cline{3-4}
        \multicolumn{3}{r|}{\multirow{2}{*}{$1$}}                        &  $0\,,\,1\,,\,2$      \\
        \multicolumn{3}{r|}{ }                                           &  $2:3:7$              \\
        \cline{4-4}
      \end{tabular}
    \end{center}
  \end{table}
Little is known about quantum gravity and in particular which symmetry is conserved by quantum gravitational interactions. We are thus led to make different assumptions on the type of processes  which are allowed by these interactions. We determined the decay branching ratios for five different models corresponding to different assumptions about which symmetry is conserved or violated by quantum gravity. In all cases, we shall assume that Lorentz symmetry is conserved.  All models take the full range of Standard Model particles as possible decay products into account and consider neutrinos to be Dirac particles which can be both left and right handed. In case of the Higgs boson, we assume that three of the four degrees of freedom merge with the degrees of freedom of the W and Z bosons. The models also allow massless gravitons as possible decay products. Furthermore, the following restrictions on the symmetries for the different models have been considered, respectively:
  \begin{enumerate}
  \item Lepton flavor, quark flavor, B-L and Lorentz invariance conserved,
  \item lepton flavor violated, everything else conserved,
  \item quark flavor violated, everything else conserved,
  \item quark and lepton flavor violated, B-L conserved,
  \item and everything violated except Lorentz invariance.
  \end{enumerate}
  

\subsubsection*{Decay of QBHs}

The branching ratios for the QBHs produced in specific reactions  for the different models and decay channels can be found in Tables (\ref{QBHuu}) to (\ref{QBHgg}). Note that the term branching ratio may be misleading as we do not use it in its usual sense, i.e. to describe the probability of the decay of a specific particle in a specific mode, but rather to describe the decomposition of a collection of black holes created by the same specific initial states into a specific mode. To determine the number of possible states, one has to consider the color multiplets and spin factors, e.g. in case of a black hole created by two up-quarks and with an incidental $\text{U}(1)_{\text{em}}$ charge of 4/3 (see i.e. Table \ref{QBHuu}), we find ${3} \times {3}= {6} + {\overline 3}$ for the color states when decaying into two up quarks. This has to be weighed by the spin factors; since both states, spin-0 as well as spin-1, are possible, we multiply the amount of states with the sum of both spin factors. One therefore 
finds:
\begin{equation}
  \textbf{QBH}(\textbf{u} , \textbf{u}, \bf{4/3}) \rightarrow \textbf{u} + \textbf{u} : \hspace{0.5cm} (6+3) \times (1+3)
\end{equation}
In case of the first two models in which quark flavor is conserved, a quantum black hole can only decay into the quarks with the initial flavor. For all other models all up-type quarks are weighed in the same way. For the models which allow quark and lepton flavor violation, we obtain alternatively:
\begin{equation}
  \textbf{QBH}(\textbf{u} , \textbf{u}, \bf{4/3}) \rightarrow \textbf{l}^{+} + \bar{\textbf{d}} : \hspace{0.5cm} (3) \times (1+3) \,.
\end{equation}
By following the same methods and considering the total number of final states, we find the branching ratios to be the quotient of the number of desired final states devided by the total number of final states. The index ``initial'' indicates the flavor of the colliding particle, whereas ``else'' sums over all other type-like particles except the initial one. The indices ``i'' and ``j'' are used to distinguish between particles of the same flavor and different flavor, i.e. $\text{i} \neq \text{j}$. Note that some of our models overlap with those considered in \cite{Gingrich:2009hj}.


\begin{table}[!ht]
  \begin{center}
    \begin{tabular}[c]{|c|c|c|c|c|c|c|}
      \hline
      \multicolumn{2}{|c|}{Final state} & \multicolumn{5}{c|}{Branching Ratios [\%]} \\
      \hline\hline
      Particle 1 & Particle 2 & Model 1 & Model 2 & Model 3 & Model 4 & Model 5  \\
      \hline
      $\text{u}_{\text{initial}}$ &  $\text{u}_{\text{initial}}$  & 100    & 100    & 16.67  & 11.11 & 11.11   \\
      $\text{u}_{\text{else}}$    &  $\text{u}_{\text{else}}$     &   0    &   0    & 83.33  & 55.56 & 55.56   \\
      $\text{l}^{+}$              &  $\bar{\text{d}}$             &   0    &   0    &   0    & 33.33 & 33.33   \\
      \hline
    \end{tabular}
    \caption{\textbf{QBH($\text{u}$ , $\text{u}$ , $4/3$)}\label{QBHuu}: Branching ratios of the decay channels of QBHs created by two up-type quarks (u). ``initial'' indicates the flavor of the incoming particles, ``else'' all other up-type quark flavors. $\text{l}^{+}$ describes an anti-lepton of any generation, $\bar{\text{d}}$ an anti-down-type quark.}
  \end{center}
\end{table}

\begin{table}[!ht]
  \begin{center}
    \begin{tabular}[c]{|c|c|c|c|c|c|c|}
      \hline
      \multicolumn{2}{|c|}{Final state} & \multicolumn{5}{c|}{Branching Ratios [\%]} \\
      \hline\hline
      Particle 1 & Particle 2 & Model 1 & Model 2 & Model 3 & Model 4 & Model 5  \\
      \hline
      $\text{d}_{\text{initial}}$ &  $\text{d}_{\text{initial}}$  & 100    & 100    & 16.67  & 8.33  &  6.67   \\
      $\text{d}_{\text{else}}$    &  $\text{d}_{\text{else}}$     &   0    &   0    & 83.33  & 41.67 & 33.33   \\
      $\text{l}^{-}$              & $\bar{\text{d}}$              &   0    &   0    &   0    & 25    & 20      \\
      $\bar{\text{u}}$            & $\nu$                         &   0    &   0    &   0    &  0    & 20      \\
      $\bar{\text{u}}$            & $\bar{\nu}$                   &   0    &   0    &   0    & 25    & 20      \\
      \hline
    \end{tabular}
    \caption{\textbf{QBH($\text{d}$ , $\text{d}$ , $-2/3$)}: Branching ratios of the decay channels of QBHs created by two down-type quarks (d). ``initial'' indicates the flavor of the incoming particles, ``else'' all other down-type quark flavors. $\text{l}^{-}$ describes a lepton of any generation, $\bar{\text{u}}$ an anti-up-type quark, and $\stackrel{(-)}{\nu}$ an (anti)neutrino.}
  \end{center}
\end{table}

\begin{table}[!ht]
  \begin{center}
    \begin{tabular}[c]{|c|c|c|c|c|c|c|}
      \hline
      \multicolumn{2}{|c|}{Final state} & \multicolumn{5}{c|}{Branching Ratios [\%]} \\
      \hline\hline
      Particle 1 & Particle 2 & Model 1 & Model 2 & Model 3 & Model 4 & Model 5  \\
      \hline
      $\text{u}_{\text{initial}}$ &  $\bar{\text{d}}_{\text{initial}}$  & 100    & 100    &   9.42 &  8.87 &  8.14   \\
      $\text{u}_{\text{else}}$    &  $\bar{\text{d}}_{\text{else}}$     &   0    &   0    &  75.39 & 70.94 & 65.16   \\
      $\text{l}^{+}_{\text{i}}$   & $\nu_{\text{i}}$                    &   0    &   0    &   3.14 &  2.96 &  2.71   \\
      $\text{l}^{+}_{\text{i}}$  & $\nu_{\text{j}}$                    &   0    &   0    &   0    &  5.91 &  5.43   \\
      $\text{l}^{+}$              	 & $\bar{\nu}$                         &   0    &   0    &   0    &  0    &  8.14   \\
      $\text{W}^{+}$              & $\text{g}$                          &   0    &   0    &   8.38 &  7.88 &  7.24   \\
      $\text{W}^{+}$              & $\gamma$                            &   0    &   0    &   1.05 &  0.99 &  0.9    \\
      $\text{W}^{+}$              & $\text{Z}^{0}$                      &   0    &   0    &   1.05 &  0.99 &  0.9    \\
      $\text{W}^{+}$              & $\text{H}$                          &   0    &   0    &   0.79 &  0.74 &  0.68   \\
      $\text{W}^{+}$              & $\text{G}$                          &   0    &   0    &   0.79 &  0.74 &  0.68   \\
      \hline
    \end{tabular}
  \caption{\textbf{QBH($\text{u}$ , $\bar{\text{d}}$ , 1)}: Branching ratios of the decay channels of QBHs created by an up- (u) and an anti-down-type quark ($\bar{\text{d}}$). ``initial'' indicates the flavor of the incoming particles, ``else'' all other (anti)quark flavors. i and j distinguish between particles of different flavors and generations, i.e. $ \text{i} \neq \text{j}$.  $\text{l}^{+}$ describes an anti-lepton of any generation, $\stackrel{(-)}{\nu}$ an (anti)neutrino.}
  \end{center}
\end{table}

\newpage

\begin{table}[!ht]
  \begin{center}
    \begin{tabular}[c]{|c|c|c|c|c|c|c|}
      \hline
      \multicolumn{2}{|c|}{Final state} & \multicolumn{5}{c|}{Branching Ratios [\%]} \\
      \hline\hline
      Particle 1 & Particle 2 & Model 1 & Model 2 & Model 3 & Model 4 & Model 5  \\
      \hline
      $\text{u}_{\text{initial}}$ &  $\text{d}_{\text{initial}}$  & 100    & 100    & 11.1   &  6.67 &  5.56   \\
      $\text{u}_{\text{else}}$    &  $\text{d}_{\text{else}}$     &   0    &   0    & 88.89  & 53.33 & 44.44   \\
      $\text{l}^{+}$              & $\bar{\text{u}}$              &   0    &   0    &   0    & 20    & 16.67  \\
      $\bar{\text{d}}$            & $\nu$                         &   0    &   0    &   0    &  0    & 16.67   \\
      $\bar{\text{d}}$            & $\bar{\nu}$                   &   0    &   0    &   0    & 20    & 16.67   \\
      \hline
    \end{tabular}
  \caption{\textbf{QBH($\text{u}$ , $\text{d}$ , $1/3$)}: Branching ratios of the decay channels of QBHs created by an up- (u) and down-type quark (d). ``initial'' indicates the flavor of the incoming particles, ``else'' all other quark flavors. $\text{l}^{+}$ describes an anti-lepton of any generation, $\stackrel{(-)}{\nu}$ an (anti)neutrino.}
  \end{center}
\end{table}

\begin{table}[!ht]
  \begin{center}
    \begin{tabular}[c]{|c|c|c|c|c|c|c|}
      \hline
      \multicolumn{2}{|c|}{Final state} & \multicolumn{5}{c|}{Branching Ratios [\%]} \\
      \hline\hline
      Particle 1 & Particle 2 & Model 1 & Model 2 & Model 3 & Model 4 & Model 5  \\
      \hline
      $\text{u}_{\text{initial}}$ &  $\text{g}$      & 72.73  & 72.73  & 22.22  & 22.22 & 22.22   \\
      $\text{u}_{\text{initial}}$ &  $\gamma$        &  9.09  &  9.09  &  2.78  &  2.78 &  2.78   \\
      $\text{u}_{\text{initial}}$ &  $\text{Z}^{0}$  &  9.09  &  9.09  &  2.78  &  2.78 &  2.78   \\
      $\text{u}_{\text{initial}}$ &  $\text{H}$      &  3.03  &  3.03  &  0.93  &  0.93 &  0.93   \\
      $\text{u}_{\text{initial}}$ &  $\text{G}$      &  6.06  &  6.06  &  1.85  &  1.85 &  1.85   \\
      $\text{u}_{\text{else}}$    &  $\text{g}$      &   0    &   0    & 44.44  & 44.44 & 44.44   \\
      $\text{u}_{\text{else}}$    &  $\gamma$        &   0    &   0    &  5.56  &  5.56 &  5.56   \\
      $\text{u}_{\text{else}}$    &  $\text{Z}^{0}$  &   0    &   0    &  5.56  &  5.56 &  5.56   \\
      $\text{u}_{\text{else}}$    &  $\text{H}$      &   0    &   0    &  1.85  &  1.85 &  1.85   \\
      $\text{u}_{\text{else}}$    &  $\text{G}$      &   0    &   0    &  3.70  &  3.70 &  3.70   \\
      $\text{d}$                  			&  $\text{W}^{+}$  &  0     &   0    &  8.33  &  8.33 &  8.33  \\
      \hline
    \end{tabular}
      \caption{\textbf{QBH($\text{u}$ , $\text{g}$ , $2/3$)}: Branching ratios of the decay channels of QBHs created by an up-type quark (u) and a gluon (g). ``initial'' indicates the flavor of the incoming particles, ``else'' all other (anti)quark flavors. $\text{d}$ describes a down-type quark.}
  \end{center}
\end{table}

\newpage
\clearpage

\begin{table}[!ht]
  \begin{center}
    \begin{tabular}[c]{|c|c|c|c|c|c|c|}
      \hline
      \multicolumn{2}{|c|}{Final state} & \multicolumn{5}{c|}{Branching Ratios [\%]} \\
      \hline\hline
      Particle 1 & Particle 2 & Model 1 & Model 2 & Model 3 & Model 4 & Model 5  \\
      \hline
		$\text{u}_{\text{i}}$ & $\bar{\text{u}}_{\text{i}}$ & 29.43 & 23.33 & 13.52 & 12.75 & 12.07\\
		$\text{u}_{\text{i}}$ & $\bar{\text{u}}_{\text{j}}$ & 0.00 & 0.00 & 27.03 & 25.50 & 24.13\\
		$\text{d}_{\text{i}}$ & $\bar{\text{d}}_{\text{i}}$ & 29.43 & 23.33 & 13.52 & 12.75 & 12.07\\
		$\text{d}_{\text{i}}$ & $\bar{\text{d}}_{\text{j}}$ & 0.00 & 0.00 & 27.03 & 25.50 & 24.13\\
		$\text{l}^{-}_{\text{i}}$ & $\text{l}^{+}_{\text{i}}$ & 3.27 & 2.59 & 1.50 & 1.42 & 1.34\\
		$\text{l}^{-}_{\text{i}}$ & $\text{l}^{+}_{\text{j}}$ & 0.00 & 5.18 & 0.00 & 2.83 & 2.68\\
		$\nu$ & $\nu$ & 0.00 & 5.18 & 0.00 & 0.00 & 2.68\\
		$\nu_{\text{i}}$ & $\bar{\nu}_{\text{i}}$ & 3.27 & 2.59 & 1.50 & 1.42 & 1.34\\
		$\nu_{\text{i}}$ & $\bar{\nu}_{\text{j}}$ & 0.00 & 5.18 & 0.00 & 2.83 & 2.68\\
		$\bar{\nu}$ & $\bar{\nu}$ & 0.00 & 5.18 & 0.00 & 0.00 & 2.68\\
		$\text{g}$ & $\text{g}$ & 2.45 & 1.94 & 1.13 & 1.06 & 1.01\\
		$\text{g}$ & $\gamma$ & 2.18 & 1.73 & 1.00 & 0.94 & 0.89\\
		$\text{g}$ & $\text{Z}^{0}$ & 8.72 & 6.91 & 4.01 & 3.78 & 3.58\\
		$\text{g}$ & $\text{H}$ & 6.54 & 5.18 & 3.00 & 2.83 & 2.68\\
		$\text{g}$ & $\text{G}$ & 6.54 & 5.18 & 3.00 & 2.83 & 2.68\\
		$\gamma$ & $\gamma$ & 0.27 & 0.22 & 0.13 & 0.12 & 0.11\\
		$\gamma$ & $\text{Z}^{0}$ & 1.09 & 0.86 & 0.50 & 0.47 & 0.45\\
		$\gamma$ & $\text{H}$ & 0.82 & 0.65 & 0.38 & 0.35 & 0.34\\
		$\gamma$ & $\text{G}$ & 0.82 & 0.65 & 0.38 & 0.35 & 0.34\\
		$\text{Z}^{0}$ & $\text{Z}^{0}$ & 1.09 & 0.86 & 0.50 & 0.47 & 0.45\\
		$\text{Z}^{0}$ & $\text{H}$ & 0.82 & 0.65 & 0.38 & 0.35 & 0.34\\
		$\text{Z}^{0}$ & $\text{G}$ & 0.82 & 0.65 & 0.38 & 0.35 & 0.34\\
		$\text{H}$ & $\text{H}$ & 0.27 & 0.22 & 0.13 & 0.12 & 0.11\\
		$\text{G}$ & $\text{G}$ & 1.09 & 0.86 & 0.50 & 0.47 & 0.45\\
		$\text{W}^{-}$ & $\text{W}^{+}$ & 1.09 & 0.86 & 0.50 & 0.47 & 0.45\\
      \hline
    \end{tabular}
      \caption{\textbf{QBH($\text{q}_{\text{i}}$ , $\bar{\text{q}}_{\text{i}}$ , $0$)} \label{QBHqqbarii}: Branching ratios of the decay channels of QBHs created by a quark-anti-quark pair of the same flavor. ``initial'' indicates the flavor of the incoming particles, ``else'' all other (anti)quark flavors. i and j distinguish between particles of different flavors and generations, i.e. $ \text{i} \neq \text{j}$.  ($\text{l}^{+}$) $\text{l}^{-}$ describes an (anti)lepton of any generation, $\stackrel{(-)}{\nu}$ an (anti)neutrino, $\text{d}$ a down-type quark, and $\text{u}$ an up-type quark.}
  \end{center}
\end{table}

\newpage

\begin{table}[!ht]
  \begin{center}
    \begin{tabular}[c]{|c|c|c|c|c|c|c|}
      \hline
      \multicolumn{2}{|c|}{Final state} & \multicolumn{5}{c|}{Branching Ratios [\%]} \\
      \hline\hline
      Particle 1 & Particle 2 & Model 1 & Model 2 & Model 3 & Model 4 & Model 5  \\
      \hline
		$\text{q}_{\text{initial}}$ & $\bar{\text{q}}_{\text{initial}}$ & 100.00 & 100.00 & 4.51 & 4.25 & 4.02\\
		$\text{q}_{\text{else}}$ & $\bar{\text{q}}_{\text{else}}$ & 0.00 & 0.00 & 76.60 & 72.26 & 68.38\\
		$\text{l}^{-}_{\text{i}}$ & $\text{l}^{+}_{\text{i}}$ & 0.00 & 0.00 & 1.50 & 1.42 & 1.34\\
		$\text{l}^{-}_{\text{i}}$ & $\text{l}^{+}_{\text{j}}$ & 0.00 & 0.00 & 0.00 & 2.83 & 2.68\\
		$\nu$ & $\nu$ & 0.00 & 0.00 & 0.00 & 0.00 & 2.68\\
		$\nu_{\text{i}}$ & $\bar{\nu}_{\text{i}}$ & 0.00 & 0.00 & 1.50 & 1.42 & 1.34\\
		$\nu_{\text{i}}$ & $\bar{\nu}_{\text{j}}$ & 0.00 & 0.00 & 0.00 & 2.83 & 2.68\\
		$\bar{\nu}$ & $\bar{\nu}$ & 0.00 & 0.00 & 0.00 & 0.00 & 2.68\\
		$\text{g}$ & $\text{g}$ & 0.00 & 0.00 & 1.13 & 1.06 & 1.01\\
		$\text{g}$ & $\gamma$ & 0.00 & 0.00 & 1.00 & 0.94 & 0.89\\
		$\text{g}$ & $\text{Z}^{0}$ & 0.00 & 0.00 & 4.01 & 3.78 & 3.58\\
		$\text{g}$ & $\text{H}$ & 0.00 & 0.00 & 3.00 & 2.83 & 2.68\\
		$\text{g}$ & $\text{G}$ & 0.00 & 0.00 & 3.00 & 2.83 & 2.68\\
		$\gamma$ & $\gamma$ & 0.00 & 0.00 & 0.13 & 0.12 & 0.11\\
		$\gamma$ & $\text{Z}^{0}$ & 0.00 & 0.00 & 0.50 & 0.47 & 0.45\\
		$\gamma$ & $\text{H}$ & 0.00 & 0.00 & 0.38 & 0.35 & 0.34\\
		$\gamma$ & $\text{G}$ & 0.00 & 0.00 & 0.38 & 0.35 & 0.34\\
		$\text{Z}^{0}$ & $\text{Z}^{0}$ & 0.00 & 0.00 & 0.50 & 0.47 & 0.45\\
		$\text{Z}^{0}$ & $\text{H}$ & 0.00 & 0.00 & 0.38 & 0.35 & 0.34\\
		$\text{Z}^{0}$ & $\text{G}$ & 0.00 & 0.00 & 0.38 & 0.35 & 0.34\\
		$\text{H}$ & $\text{H}$ & 0.00 & 0.00 & 0.13 & 0.12 & 0.11\\
		$\text{G}$ & $\text{G}$ & 0.00 & 0.00 & 0.50 & 0.47 & 0.45\\
		$\text{W}^{-}$ & $\text{W}^{+}$ & 0.00 & 0.00 & 0.50 & 0.47 & 0.45\\
      \hline
    \end{tabular}
    \caption{\textbf{QBH($\text{q}_{\text{i}}$ , $\bar{\text{q}}_{\text{j}}$ , $0$)}\label{QBHqqbarij}: Branching ratios of the decay channels of QBHs created by a quark-anti-quark pair of different flavor. ``initial'' indicates the flavor of the incoming particles, ``else'' all other (anti)quark flavors. i and j distinguish between particles of different flavors and generations, i.e. $ \text{i} \neq \text{j}$.  ($\text{l}^{+}$) $\text{l}^{-}$ describes an (anti)lepton of any generation, $\stackrel{(-)}{\nu}$ an (anti)neutrino, $\text{d}$ a down-type quark, and $\text{u}$ an up-type quark.}
  \end{center}
\end{table}

\newpage

\begin{table}[!ht]
  \begin{center}
    \begin{tabular}[c]{|c|c|c|c|c|c|c|}
      \hline
      \multicolumn{2}{|c|}{Final state} & \multicolumn{5}{c|}{Branching Ratios [\%]} \\
      \hline\hline
      Particle 1 & Particle 2 & Model 1 & Model 2 & Model 3 & Model 4 & Model 5  \\
      \hline
		$\text{q}_{\text{i}}$ & $\bar{\text{q}}_{\text{i}}$ & 17.65 & 17.16 & 13.17 & 12.97 & 12.77\\
		$\text{q}_{\text{i}}$ & $\bar{\text{q}}_{\text{j}}$ & 0.00 & 0.00 & 26.34 & 25.94 & 25.55\\
		$\text{l}^{-}_{\text{i}}$ & $\text{l}^{+}_{\text{i}}$ & 1.04 & 0.50 & 0.39 & 0.38 & 0.38\\
		$\text{l}^{-}_{\text{i}}$ & $\text{l}^{+}_{\text{j}}$ & 0.00 & 1.01 & 0.00 & 0.76 & 0.75\\
		$\nu$ & $\nu$ & 0.00 & 0.84 & 0.00 & 0.00 & 0.63\\
		$\nu_{\text{i}}$ & $\bar{\nu}_{\text{i}}$ & 1.30 & 0.50 & 0.39 & 0.38 & 0.38\\
		$\nu_{\text{i}}$ & $\bar{\nu}_{\text{j}}$ & 0.00 & 1.01 & 0.00 & 0.76 & 0.75\\
		$\bar{\nu}$ & $\bar{\nu}$ & 0.00 & 1.01 & 0.00 & 0.00 & 0.75\\
		$\text{g}$ & $\text{g}$ & 49.83 & 48.44 & 37.19 & 36.62 & 36.07\\
		$\text{g}$ & $\gamma$ & 12.46 & 12.11 & 9.30 & 9.15 & 9.02\\
		$\text{g}$ & $\text{Z}^{0}$ & 12.46 & 12.11 & 9.30 & 9.15 & 9.02\\
		$\gamma$ & $\gamma$ & 0.78 & 0.76 & 0.58 & 0.57 & 0.56\\
		$\gamma$ & $\text{Z}^{0}$ & 0.78 & 0.76 & 0.58 & 0.57 & 0.56\\
		$\text{Z}^{0}$ & $\text{Z}^{0}$ & 0.78 & 0.76 & 0.58 & 0.57 & 0.56\\
		$\text{Z}^{0}$ & $\text{H}$ & 0.00 & 0.00 & 0.00 & 0.00 & 0.00\\
		$\text{Z}^{0}$ & $\text{G}$ & 0.61 & 0.59 & 0.45 & 0.45 & 0.44\\
		$\text{H}$ & $\text{G}$ & 0.17 & 0.17 & 0.13 & 0.13 & 0.13\\
		$\text{H}$ & $\text{H}$ & 0.61 & 0.59 & 0.45 & 0.45 & 0.44\\
		$\text{G}$ & $\text{G}$ & 0.78 & 0.76 & 0.58 & 0.57 & 0.56\\
		$\text{W}^{-}$ & $\text{W}^{+}$ & 0.78 & 0.76 & 0.58 & 0.57 & 0.56\\
      \hline
    \end{tabular}
      \caption{\textbf{QBH($\text{g}$, $\text{g}$ , $0$)} \label{QBHgg}: Branching ratios of the decay channels of QBHs created by two gluons. ``initial'' indicates the flavor of the incoming particles, ``else'' all other (anti)quark flavors. i and j distinguish between particles of different flavors and generations, i.e. $ \text{i} \neq \text{j}$.  ($\text{l}^{+}$) $\text{l}^{-}$ describes an (anti)lepton of any generation, $\stackrel{(-)}{\nu}$ an (anti)neutrino, $\text{d}$ a down-type quark, and $\text{u}$ an up-type quark.}
  \end{center}
\end{table}

\newpage
\clearpage

As aforementioned, we assume that there are five quantum black hole states between the Planck mass and the continuous semi-classical region. We thus calculate the cross section for these five states for different models with low scale quantum gravity. Our results can be found in Tables (\ref{T1}) to (\ref{T3}) which correspond to different center of mass energies (7, 8 and 14 TeV). We emphasize that these intermediate black holes will have different spins. We are summing over these allowed spin states in the cross sections given in Tables (\ref{T1}) to (\ref{T3}).

The cross section for a specific final state, e.g. $\text{QBH} \rightarrow \gamma + \gamma$, is obtained by summing over the production cross sections of the contributing QBHs  which have been multiplied with the desired branching ratio (cf. Table (\ref{QBHuu}) to (\ref{QBHgg})):
\begin{equation*}
	\sigma_{\text{QBH} \rightarrow \text{final state}} = \sum_{\stackrel{\text{\tiny{average}}}{\text{\tiny{initial states}}}}\, \sum_{\text{\tiny{masses}}} \, \sigma_{\text{initial state} \rightarrow \text{QBH}_\text{mass}} \times \text{BR}_{\text{final state}, \text{model}} \, .
\end{equation*}
Each cross section for a specific initial states was calculated via equations (\ref{pp_total}) and (\ref{pp_individual}) by using a Monte Carlo integration algorithm. In the case of two photon production, the relevant initial states are a quark-anti-quark pair and two gluons. Assuming a specific model, we sum up the cross sections for all contributing masses up to the CoM energy and then multiply the result with the suitable branching ratios (Table \ref{QBHqqbarii}, \ref{QBHqqbarij}, and  \ref{QBHgg}). The desired cross section for the chosen final state is given by taking the average over all initial state contributions.


\begin{table}[!ht]
\begin{center}
\begin{tabular}{|c|c|c|c|c|c|}\hline
cross section in fb & \multicolumn{5}{c|}{Model for low scale quantum gravity}\\\hline
$\sigma \left( p + p \rightarrow \text{any QBH} \right)$ & 4 dim\cite{Calmet:2008tn}  & ADD n = 5 & ADD n = 6 & ADD n = 7 & RS\\\hline\hline
\multicolumn{6}{|c|}{CoM energy of $\sqrt{\text{s}} = 7\ \text{TeV}$}\\\hline\hline
\multicolumn{6}{|c|}{$\text{M}_{\text{D}} = 1\ \text{TeV}$}\\\hline
$ \text{M}_{\text{BH}} = 1\ \text{TeV} $ 
& $2.21 \times 10^{4}$ & $1.54 \times 10^{7}$ & $2.09 \times 10^{7}$ & $2.66 \times 10^{7}$ & $7.42 \times 10^{5}$\\
\hline
$ \text{M}_{\text{BH}} = 2\ \text{TeV} $ 
& $4.52 \times 10^{3}$ & $1.25 \times 10^{5}$ & $1.30 \times 10^{6}$ & $1.62 \times 10^{6}$ & $7.59 \times 10^{4}$ \\
\hline
$ \text{M}_{\text{BH}} = 3\ \text{TeV} $ 
& 551.90  	      & $8.90 \times 10^{4}$ & $7.91 \times 10^{4}$  & $9.71 \times 10^{4}$  & $6.16 \times 10^{3}$ \\
\hline
$ \text{M}_{\text{BH}} = 4\ \text{TeV} $ 
& 30.81		      & $3.38 \times 10^{3}$ & $2.70 \times 10^{3}$  & $3.28 \times 10^{3}$  & 258.13 \\
\hline
$ \text{M}_{\text{BH}} = 5\ \text{TeV} $ 
& 0.45  	      & 21.31		     & 26.66		     & 32.15 		     & 2.99 \\
\hline\hline
\multicolumn{6}{|c|}{$\text{M}_{\text{D}} = 3\ \text{TeV}$}\\\hline
$ \text{M}_{\text{BH}} = 3\ \text{TeV} $ 
& 5.14  & $3.56 \times 10^{3}$	& $4.85 \times 10^{3}$	& $6.19 \times 10^{3}$	& 172.40 \\
\hline
$ \text{M}_{\text{BH}} = 4\ \text{TeV} $ 
& 0.27  & 115.77		& 154.20		& 194.95		& 6.74   \\
\hline
$ \text{M}_{\text{BH}} = 5\ \text{TeV} $ 
& 0.004 & 1.06			& 1.40			& 1.75	 		& 0.07   \\
\hline
$ \text{M}_{\text{BH}} = 6\ \text{TeV} $ 
& $1.97 \times 10^{-6}$     	& $4.34 \times 10^{-4}$	 & 0.001			& 0.001 		& $3.31 \times 10^{-5}$    \\
\hline\hline
\multicolumn{6}{|c|}{$\text{M}_{\text{D}} = 5\ \text{TeV}$}\\\hline
$ \text{M}_{\text{BH}} = 5\ \text{TeV} $ 
& $3.83 \times 10^{-4}$  & 0.27 & 0.36 & 0.46 & 0.01	 \\
\hline
$ \text{M}_{\text{BH}} = 6\ \text{TeV} $ 
& $2.01 \times 10^{-7}$  & $1.03 \times 10^{-4}$   & $1.38 \times 10^{-4}$ 	  & $1.76 \times 10^{-4}$	 & $5.61 \times 10^{-6}$	 \\
\hline
\end{tabular}
\caption{Total cross sections for the production of QBHs with masses above the chosen reduced Planck mass of $\text{M}_{\text{D}} = 1, 3 \ \text{and} \ 5 \ \text{TeV}$, respectively, and below CoM energy of the initial particles of 7 TeV. \label{T1}
}
\end{center}
\end{table}

\begin{table}[!ht]
\begin{center}
\begin{tabular}{|c|c|c|c|c|c|}\hline
cross section in fb & \multicolumn{5}{c|}{Model for low scale quantum gravity}\\\hline
$\sigma \left( p + p \rightarrow \text{any QBH} \right)$ & 4 dim\cite{Calmet:2008tn} & ADD n = 5 & ADD n = 6 & ADD n = 7 & RS\\\hline\hline
\multicolumn{6}{|c|}{CoM energy of $\sqrt{\text{s}} = 8\ \text{TeV}$}\\\hline\hline
\multicolumn{6}{|c|}{$\text{M}_{\text{D}} = 1\ \text{TeV}$}\\\hline
$ \text{M}_{\text{BH}} = 1\ \text{TeV} $ 
& $3.44 \times 10^{5}$ 	& $2.40 \times 10^{7}$ 	& $3.24 \times 10^{7}$ 	& $4.14 \times 10^{7}$ 	& $1.15 \times 10^{6}$ \\
\hline
$ \text{M}_{\text{BH}} = 2\ \text{TeV} $ 
& $9.16 \times 10^{3}$ 	& $2.01 \times 10^{6}$ 	& $2.63 \times 10^{6}$ 	& $3.28 \times 10^{6}$ 	& $1.53 \times 10^{5}$ \\
\hline
$ \text{M}_{\text{BH}} = 3\ \text{TeV} $ 
& $1.71 \times 10^{3}$ 	& $1.91 \times 10^{5}$ 	& $2.45 \times 10^{5}$ 	& $3.01 \times 10^{5}$ 	& $1.91 \times 10^{7}$ \\
\hline
$ \text{M}_{\text{BH}} = 4\ \text{TeV} $ 
& 192.29	  	& $1.33 \times 10^{4}$ 	& $1.68 \times 10^{4}$ 	& $2.04 \times 10^{4}$ 	& $1.61 \times 10^{6}$ \\
\hline
$ \text{M}_{\text{BH}} = 5\ \text{TeV} $ 
& 10.13 		& 483.33		& 604.51		& 729.04		& 67.90			\\
\hline\hline
\multicolumn{6}{|c|}{$\text{M}_{\text{D}} = 3\ \text{TeV}$}\\\hline
$ \text{M}_{\text{BH}} = 3\ \text{TeV} $ 
& 16.31	& $1.14 \times 10^{4}$	& $1.54 \times 10^{4}$ 	& $1.96 \times 10^{4}$ 	& 546.61 \\
\hline
$ \text{M}_{\text{BH}} = 4\ \text{TeV} $ 
& 1.73	& 748.93		& 998.55		& $1.26 \times 10^{3}$ 	& 43.59 \\
\hline
$ \text{M}_{\text{BH}} = 5\ \text{TeV} $ 
& 0.09	& 25.47			& 33.57			& 42.11 		& 1.72 \\
\hline
$ \text{M}_{\text{BH}} = 6\ \text{TeV} $ 
& $1.10 \times 10^{-3}$	& 0.24 			& 0.32			& 0.39 			& 0.02 \\
\hline
$ \text{M}_{\text{BH}} = 7\ \text{TeV} $ 
& $6.19 \times 10^{-7}$ & $1.05 \times 10^{-4}$ & $1.36 \times 10^{-4}$ & $1.69 \times 10^{-4}$ & $8.89 \times 10^{-6}$ \\
\hline\hline
\multicolumn{6}{|c|}{$\text{M}_{\text{D}} = 5\ \text{TeV}$}\\\hline
$ \text{M}_{\text{BH}} = 5\ \text{TeV} $ 
& 0.01	& 6.58	& 8.88	& 11.34	& 0.32	 \\
\hline
$ \text{M}_{\text{BH}} = 6\ \text{TeV} $ 
& $ 1.17 \times 10^{-4}$	& 0.06	& 0.08	& 0.10	& $ \times 10^{-7}$	 \\
\hline
$ \text{M}_{\text{BH}} = 7\ \text{TeV} $ 
& $6.24 \times 10^{-8}$	& $2.48 \times 10^{-5}$	& $3.30 \times 10^{-5}$	& $4.17 \times 10^{-5}$	& $1.49 \times 10^{-6}$	 \\
\hline
\end{tabular}
\caption{Total cross sections for the production of QBHs with masses above the chosen reduced Planck mass of $\text{M}_{\text{D}} = 1, 3 \ \text{and} \ 5 \ \text{TeV}$, respectively, and below CoM energy of the initial particles of 8 TeV.\label{T2}}
\end{center}
\end{table}

\begin{table}[!ht]
\begin{center}
\begin{tabular}{|c|c|c|c|c|c|}\hline
cross section in fb & \multicolumn{5}{c|}{Model for low scale quantum gravity}\\\hline
$\sigma \left( p + p \rightarrow \text{any QBH} \right)$ & 4 dim\cite{Calmet:2008tn}  & ADD n = 5 & ADD n = 6 & ADD n = 7 & RS\\\hline\hline
\multicolumn{6}{|c|}{CoM energy of $\sqrt{\text{s}} = 14\ \text{TeV}$}\\\hline\hline
\multicolumn{6}{|c|}{$\text{M}_{\text{D}} = 1\ \text{TeV}$}\\\hline
$ \text{M}_{\text{BH}} = 1 \ \text{TeV} $ 
& $1.72 \times 10^{5}$ & $1.20 \times 10^{8}$ & $1.62 \times 10^{8}$ & $2.07 \times 10^{8}$ & $5.77 \times 10^{6}$ \\
\hline
$ \text{M}_{\text{BH}} = 2 \ \text{TeV} $ 
& $8.86 \times 10^{4}$ & $1.95 \times 10^{7}$ & $2.54 \times 10^{7}$ & $3.17 \times 10^{7}$ & $1.48 \times 10^{6}$ \\
\hline
$ \text{M}_{\text{BH}} = 3 \ \text{TeV} $ 
& $4.22 \times 10^{4}$ & $4.71 \times 10^{6}$ & $6.04 \times 10^{6}$ & $7.42 \times 10^{6}$ & $4.71 \times 10^{5}$ \\
\hline
$ \text{M}_{\text{BH}} = 4\ \text{TeV} $ 
& $1.81 \times 10^{4}$ & $1.25 \times 10^{6}$ & $1.58 \times 10^{6}$ & $1.93 \times 10^{6}$ & $1.52 \times 10^{5}$ \\
\hline
$ \text{M}_{\text{BH}} = 5 \ \text{TeV} $ 
& $6.84 \times 10^{3}$ & $3.26 \times 10^{5}$ & $4.08 \times 10^{5}$ & $4.92 \times 10^{5}$ & $4.58 \times 10^{4}$ \\
\hline\hline
\multicolumn{6}{|c|}{$\text{M}_{\text{D}} = 3\ \text{TeV}$}\\\hline
$ \text{M}_{\text{BH}} = 3\ \text{TeV} $ 
& 430.51 & $3.00 \times 10^{5}$ & $4.05 \times 10^{5}$ & $5.18 \times 10^{5}$ & $1.44 \times 10^{4}$ \\
\hline
$ \text{M}_{\text{BH}} = 4\ \text{TeV} $ 
& 179.42 & $7.75 \times 10^{4}$ & $1.03 \times 10^{5}$ & $1.30 \times 10^{5}$ & $4.51 \times 10^{3}$ \\
\hline
$ \text{M}_{\text{BH}} = 5\ \text{TeV} $ 
& 65.73  & $1.96 \times 10^{4}$ & $2.58 \times 10^{4}$ & $3.23 \times 10^{4}$ & $1.32 \times 10^{3}$ \\
\hline
$ \text{M}_{\text{BH}} = 6\ \text{TeV} $ 
& 20.58  & $4.52 \times 10^{3}$ & $5.91 \times 10^{3}$ & $7.36 \times 10^{3}$ & 344.80 \\
\hline
$ \text{M}_{\text{BH}} = 7\ \text{TeV} $ 
& 5.31   & 902.51		& $1.17 \times 10^{3}$ & $1.45 \times 10^{3}$ & 76.28 \\
\hline\hline
\multicolumn{6}{|c|}{$\text{M}_{\text{D}} = 5 \ \text{TeV}$}\\\hline
$ \text{M}_{\text{BH}} = 5\ \text{TeV} $ 
& 7.66	& $5.34 \times 10^{3}$	& $7.21 \times 10^{3}$ & $9.21 \times 10^{3}$	& 256.53 \\
\hline
$ \text{M}_{\text{BH}} = 6\ \text{TeV} $ 
& 2.37	& $1.22 \times 10^{3}$	& $1.63 \times 10^{3}$ & $2.07 \times 10^{3}$ 	& 66.05 \\
\hline
$ \text{M}_{\text{BH}} = 7\ \text{TeV} $ 
& 0.60	& 239.64		& 318.45		& 401.91	 	& 14.41 \\
\hline
$ \text{M}_{\text{BH}} = 8\ \text{TeV} $ 
& 0.12	& 38.14			& 50.36			& 63.26			& 2.51  \\
\hline
$ \text{M}_{\text{BH}} = 9 \ \text{TeV} $ 
& 0.02	& 4.48			& 5.88			& 7.36			& 0.32  \\
\hline
\end{tabular}
\caption{Total cross sections for the production of QBHs with masses above the chosen reduced Planck mass of $\text{M}_{\text{D}} = 1, 3 \ \text{and} \ 5 \ \text{TeV}$, respectively, and below CoM energy of the initial particles of 14 TeV.  \label{T3}
}
\end{center}
\end{table}

\newpage

\begin{table}[!ht]
\begin{center}
\begin{tabular}{|c|c|c|c|c|c|}\hline
cross section in fb & \multicolumn{5}{c|}{Model for low scale quantum gravity}\\\hline
$\sigma \left(\text{QBH} \rightarrow \gamma + \gamma \right)$ & 4 dim\cite{Calmet:2008tn}  & ADD n = 5 & ADD n = 6 & ADD n = 7 & RS\\\hline\hline
\multicolumn{6}{|c|}{CoM energy of $\sqrt{\text{s}} = 7\ \text{TeV}$}\\\hline\hline
\multicolumn{6}{|c|}{$\text{M}_{\text{D}} = 1\ \text{TeV}$}\\\hline
Model 1 & 0.17 	& 109.93	& 148.32	& 189.34	& 5.36	\\
\hline
Model 2 & 0.14 	&  93.18	& 125.72	& 160.50	& 4.54	\\
\hline
Model 3 & 0.11	&  71.40	&  96.33	& 122.98	& 3.48	\\
\hline
Model 4	& 0.10  &  68.39	&  92.27	& 117.79	& 3.33	\\
\hline
Model 5	& 0.10 	&  65.67	&  88.60	& 113.11	& 3.20	\\
\hline\hline
\multicolumn{6}{|c|}{$\text{M}_{\text{D}} = 3\ \text{TeV}$}\\\hline
Model 1 & $5.26 \times 10^{-6}$	& $3.64 \times 10^{-3}$	& $4.92 \times 10^{-3}$	& $6.28 \times 10^{-3}$	& $1.76 \times 10^{-4}$	\\
\hline
Model 2 & $4.27 \times 10^{-6}$	& $2.96 \times 10^{-3}$ & $3.99 \times 10^{-3}$	& $5.10 \times 10^{-2}$	& $1.43 \times 10^{-4}$	\\
\hline
Model 3 & $3.37 \times 10^{-6}$	& $2.33 \times 10^{-3}$ & $3.15 \times 10^{-3}$	& $4.02 \times 10^{-3}$	& $1.12 \times 10^{-4}$	\\
\hline
Model 4	& $3.20 \times 10^{-6}$	& $2.21 \times 10^{-3}$ & $2.99 \times 10^{-3}$	& $3.82 \times 10^{-3}$	& $1.07 \times 10^{-4}$	\\
\hline
Model 5	& $3.04 \times 10^{-6}$	& $2.11 \times 10^{-3}$ & $2.84 \times 10^{-3}$	& $3.63 \times 10^{-3}$	& $1.01 \times 10^{-4}$	\\
\hline\hline
\multicolumn{6}{|c|}{$\text{M}_{\text{D}} = 5\ \text{TeV}$}\\\hline
Model 1 & $1.09 \times 10^{-10}$& $7.60 \times 10^{-8}$	& $1.03 \times 10^{-7}$	& $1.31 \times 10^{-7}$	& $3.65 \times 10^{-9}$	\\
\hline
Model 2 & $8.68 \times 10^{-11}$& $6.05 \times 10^{-8}$ & $8.17 \times 10^{-8}$	& $1.04 \times 10^{-7}$	& $2.91 \times 10^{-9}$	\\
\hline
Model 3 & $8.86 \times 10^{-11}$& $6.18 \times 10^{-8}$ & $8.34 \times 10^{-8}$	& $1.07 \times 10^{-7}$	& $2.97 \times 10^{-9}$	\\
\hline
Model 4	& $8.36 \times 10^{-11}$& $5.83 \times 10^{-8}$ & $7.87 \times 10^{-8}$	& $1.01 \times 10^{-7}$	& $2.80 \times 10^{-9}$	\\
\hline
Model 5	& $7.92 \times 10^{-11}$& $5.52 \times 10^{-8}$ & $7.46 \times 10^{-8}$	& $9.52 \times 10^{-8}$	& $2.65 \times 10^{-9}$	\\
\hline
\end{tabular}
\caption{Cross sections for production of a QBH decaying into two photons with a CoM energy of the initial particles of 7 TeV.\label{GAMGAM1}
}
\end{center}
\end{table}

\begin{table}[!ht]
\begin{center}
\begin{tabular}{|c|c|c|c|c|c|}\hline
cross section in fb & \multicolumn{5}{c|}{Model for low scale quantum gravity}\\\hline
$\sigma \left(\text{QBH} \rightarrow \gamma + \gamma \right)$ & 4 dim\cite{Calmet:2008tn} & ADD n = 5 & ADD n = 6 & ADD n = 7 & RS\\\hline\hline
\multicolumn{6}{|c|}{CoM energy of $\sqrt{\text{s}} = 8\ \text{TeV}$}\\\hline\hline
\multicolumn{6}{|c|}{$\text{M}_{\text{D}} = 1\ \text{TeV}$}\\\hline
Model 1 & 0.30 	& 192.12	& 259.10	& 330.65	& 9.44	\\
\hline
Model 2 & 0.25 	& 163.80	& 220.92	& 281.94	& 8.05	\\
\hline
Model 3 & 0.20	& 125.95	& 169.87	& 216.79	& 6.19\\
\hline
Model 4	& 0.19  & 120.78	& 162.90	& 207.90	& 5.93\\
\hline
Model 5	& 0.18 	& 116.11	& 156.60	& 199.86	&  5.70	\\
\hline\hline
\multicolumn{6}{|c|}{$\text{M}_{\text{D}} = 3\ \text{TeV}$}\\\hline
Model 1 & $2.47 \times 10^{-5}$	& $1.70 \times 10^{-2}$	& $2.29 \times 10^{-2}$	& $2.92 \times 10^{-2}$	& $8.19 \times 10^{-4}$	\\
\hline
Model 2 & $2.01 \times 10^{-5}$	& $1.38 \times 10^{-2}$ & $1.86 \times 10^{-2}$	& $2.38 \times 10^{-2}$	& $6.68 \times 10^{-4}$	\\
\hline
Model 3 & $1.57 \times 10^{-5}$	& $1.07 \times 10^{-2}$ & $1.45 \times 10^{-2}$	& $1.85 \times 10^{-3}$	& $5.18 \times 10^{-4}$	\\
\hline
Model 4	& $1.49 \times 10^{-5}$	& $1.02 \times 10^{-2}$ & $1.37 \times 10^{-2}$	& $1.75 \times 10^{-3}$	& $4.92 \times 10^{-4}$	\\
\hline
Model 5	& $1.41 \times 10^{-5}$	& $9.69 \times 10^{-3}$ & $1.31 \times 10^{-2}$	& $1.67 \times 10^{-3}$	& $4.68 \times 10^{-4}$	\\
\hline\hline
\multicolumn{6}{|c|}{$\text{M}_{\text{D}} = 5\ \text{TeV}$}\\\hline
Model 1 & $3.02 \times 10^{-9}$ & $2.10 \times 10^{-6}$	& $2.84 \times 10^{-6}$	& $3.62 \times 10^{-6}$	& $1.01 \times 10^{-7}$	\\
\hline
Model 2 & $2.42 \times 10^{-9}$ & $1.68 \times 10^{-6}$ & $2.27 \times 10^{-6}$	& $2.90 \times 10^{-6}$	& $8.10 \times 10^{-8}$	\\
\hline
Model 3 & $2.26 \times 10^{-9}$ & $1.57 \times 10^{-6}$ & $2.12 \times 10^{-6}$	& $2.70 \times 10^{-6}$	& $7.54 \times 10^{-8}$	\\
\hline
Model 4	& $2.13 \times 10^{-9}$ & $1.48 \times 10^{-6}$ & $2.00 \times 10^{-6}$	& $2.55 \times 10^{-6}$	& $7.13 \times 10^{-8}$	\\
\hline
Model 5	& $2.02 \times 10^{-9}$ & $1.40 \times 10^{-6}$ & $1.90 \times 10^{-6}$	& $2.42 \times 10^{-6}$	& $6.76 \times 10^{-8}$	\\
\hline
\end{tabular}
\caption{Cross sections for production of a QBH decaying into two photons with a CoM energy of the initial particles of 8 TeV.\label{GAMGAM2}}
\end{center}
\end{table}

\begin{table}[!ht]
\begin{center}
\begin{tabular}{|c|c|c|c|c|c|}\hline
cross section in fb & \multicolumn{5}{c|}{Model for low scale quantum gravity}\\\hline
$\sigma \left(\text{QBH} \rightarrow \gamma + \gamma \right)$ & 4 dim\cite{Calmet:2008tn}  & ADD n = 5 & ADD n = 6 & ADD n = 7 & RS\\\hline\hline
\multicolumn{6}{|c|}{CoM energy of $\sqrt{\text{s}} = 14\ \text{TeV}$}\\\hline\hline
\multicolumn{6}{|c|}{$\text{M}_{\text{D}} = 1\ \text{TeV}$}\\\hline
Model 1 & 2.65 	& $1.41 \times 10^{3}$ & $1.89 \times 10^{3}$	& $2.41 \times 10^{3}$	& 72.83	\\
\hline
Model 2 & 2.30 	& $1.23 \times 10^{3}$ & $1.65 \times 10^{3}$	& $2.11 \times 10^{3}$	& 63.46	\\
\hline
Model 3 & 1.78	& $9.58 \times 10^{2}$ & $1.29 \times 10^{3}$	& $1.64 \times 10^{3}$	& 43.36	\\
\hline
Model 4	& 1.71  & $9.23 \times 10^{2}$ & $1.24 \times 10^{3}$	& $1.58 \times 10^{3}$	& 47.55\\
\hline
Model 5	& 1.65 	& $8.91 \times 10^{2}$ & $1.20 \times 10^{3}$	& $1.53 \times 10^{3}$	& 45.90	\\
\hline\hline
\multicolumn{6}{|c|}{$\text{M}_{\text{D}} = 3\ \text{TeV}$}\\\hline
Model 1 & $2.47 \times 10^{-3}$	& 1.54	& 2.07	& 2.64	& $7.68 \times 10^{-2}$	\\
\hline
Model 2 & $2.04 \times 10^{-3}$	& 1.27	& 1.72	& 2.19	& $6.36 \times 10^{-2}$	\\
\hline
Model 3 & $1.56 \times 10^{-3}$	& 0.97	& 1.31	& 1.67	& $4.86 \times 10^{-2}$	\\
\hline
Model 4	& $1.49 \times 10^{-3}$	& 0.93	& 1.25	& 1.59	& $4.63 \times 10^{-2}$	\\
\hline
Model 5	& $1.42 \times 10^{-3}$	& 0.89	& 1.19	& 1.52	& $4.42 \times 10^{-2}$	\\
\hline\hline
\multicolumn{6}{|c|}{$\text{M}_{\text{D}} = 5\ \text{TeV}$}\\\hline
Model 1 & $1.51 \times 10^{-5}$ & $1.00 \times 10^{-2}$	& $1.35\times 10^{-2}$	& $1.72 \times 10^{-2}$	& $4.90 \times 10^{-4}$	\\
\hline
Model 2 & $1.23 \times 10^{-5}$ & $8.15 \times 10^{-3}$ & $1.10 \times 10^{-2}$	& $1.40 \times 10^{-2}$	& $3.99 \times 10^{-4}$	\\
\hline
Model 3 & $9.59 \times 10^{-6}$ & $6.33 \times 10^{-3}$ & $8.53 \times 10^{-3}$	& $1.09 \times 10^{-2}$	& $3.10 \times 10^{-4}$	\\
\hline
Model 4	& $9.10 \times 10^{-6}$ & $6.01 \times 10^{-3}$ & $8.10 \times 10^{-3}$	& $1.03 \times 10^{-2}$	& $2.94 \times 10^{-4}$	\\
\hline
Model 5	& $8.66 \times 10^{-6}$ & $5.72 \times 10^{-3}$ & $7.71 \times 10^{-3}$	& $9.84 \times 10^{-3}$	& $2.80 \times 10^{-4}$	\\
\hline
\end{tabular}
\caption{Cross sections for production of a QBH decaying into two photons with a CoM energy of the initial particles of 14 TeV.\label{GAMGAM3}
}
\end{center}
\end{table}

\newpage

\begin{table}[!ht]
\begin{center}
\begin{tabular}{|c|c|c|c|c|c|}\hline
cross section in fb & \multicolumn{5}{c|}{Model for low scale quantum gravity}\\\hline
$\sigma \left(\text{QBH} \rightarrow \text{e}^{+} + \text{e}^{-} \right)$ & 4 dim\cite{Calmet:2008tn}  & ADD n = 5 & ADD n = 6 & ADD n = 7 & RS\\\hline\hline
\multicolumn{6}{|c|}{CoM energy of $\sqrt{\text{s}} = 7\ \text{TeV}$}\\\hline\hline
\multicolumn{6}{|c|}{$\text{M}_{\text{D}} = 1\ \text{TeV}$}\\\hline
Model 1 & 0.24 	& 158.26	& 213.46	& 272.44	& 7.76	\\
\hline
Model 2 & 0.20 	& 126.84	& 171.09	& 218.36	& 6.22	\\
\hline
Model 3 & 0.18	& 119.93	& 161.77	& 206.48	& 5.87	\\
\hline
Model 4	& 0.17  & 113.37	& 152.92	& 195.18	& 5.55	\\
\hline
Model 5	& 0.17 	& 107.50	& 145.01	& 185.09	& 5.26	\\
\hline\hline
\multicolumn{6}{|c|}{$\text{M}_{\text{D}} = 3\ \text{TeV}$}\\\hline
Model 1 & $1.27 \times 10^{-5}$	& $8.76 \times 10^{-3}$	& $1.18 \times 10^{-2}$	& $1.51 \times 10^{-2}$	& $4.22 \times 10^{-4}$	\\
\hline
Model 2 & $1.01 \times 10^{-5}$	& $6.96 \times 10^{-3}$ & $9.40 \times 10^{-3}$	& $1.20 \times 10^{-2}$	& $3.35 \times 10^{-4}$	\\
\hline
Model 3 & $9.04 \times 10^{-6}$	& $6.24 \times 10^{-3}$ & $8.43 \times 10^{-3}$	& $1.08 \times 10^{-2}$	& $3.01 \times 10^{-4}$	\\
\hline
Model 4	& $8.53 \times 10^{-6}$	& $5.89 \times 10^{-3}$ & $7.95 \times 10^{-3}$	& $1.02 \times 10^{-2}$	& $2.84 \times 10^{-4}$	\\
\hline
Model 5	& $8.08 \times 10^{-6}$	& $5.58 \times 10^{-3}$ & $7.53 \times 10^{-3}$	& $9.62 \times 10^{-3}$	& $2.69 \times 10^{-4}$	\\
\hline\hline
\multicolumn{6}{|c|}{$\text{M}_{\text{D}} = 5\ \text{TeV}$}\\\hline
Model 1 & $3.21 \times 10^{-10}$& $2.24 \times 10^{-7}$	& $3.02 \times 10^{-7}$	& $3.86 \times 10^{-7}$	& $1.08 \times 10^{-8}$	\\
\hline
Model 2 & $2.55 \times 10^{-10}$& $1.77 \times 10^{-7}$ & $2.40 \times 10^{-7}$	& $3.06 \times 10^{-7}$	& $8.53 \times 10^{-9}$	\\
\hline
Model 3 & $2.99 \times 10^{-10}$& $2.09 \times 10^{-7}$ & $2.82 \times 10^{-7}$	& $3.60 \times 10^{-7}$	& $1.00 \times 10^{-8}$	\\
\hline
Model 4	& $2.82 \times 10^{-10}$& $1.97 \times 10^{-7}$ & $2.66 \times 10^{-7}$	& $3.39 \times 10^{-7}$	& $9.45 \times 10^{-9}$	\\
\hline
Model 5	& $2.67 \times 10^{-10}$& $1.86 \times 10^{-7}$ & $2.51 \times 10^{-7}$	& $3.21 \times 10^{-7}$	& $8.95 \times 10^{-9}$	\\
\hline
\end{tabular}
\caption{Cross sections for production of a QBH decaying into an electron and positron with a CoM energy of the initial particles of 7 TeV.\label{EE1}
}
\end{center}
\end{table}

\begin{table}[!ht]
\begin{center}
\begin{tabular}{|c|c|c|c|c|c|}\hline
cross section in fb & \multicolumn{5}{c|}{Model for low scale quantum gravity}\\\hline
$\sigma \left(\text{QBH} \rightarrow \text{e}^{+} + \text{e}^{-} \right)$ & 4 dim\cite{Calmet:2008tn} & ADD n = 5 & ADD n = 6 & ADD n = 7 & RS\\\hline\hline
\multicolumn{6}{|c|}{CoM energy of $\sqrt{\text{s}} = 8\ \text{TeV}$}\\\hline\hline
\multicolumn{6}{|c|}{$\text{M}_{\text{D}} = 1\ \text{TeV}$}\\\hline
Model 1 & 0.42 	& 263.92	& 355.78	& 453.90	& 13.07	\\
\hline
Model 2 & 0.34 	& 211.85	& 285.59	& 364.35	& 10.49	\\
\hline
Model 3 & 0.32	& 202.69	& 273.27	& 348.67	& 10.02	\\
\hline
Model 4	& 0.30  & 191.65	& 258.39	& 329.67	&  9.47	\\
\hline
Model 5	& 0.29 	& 181.78	& 245.08	& 312.69	&  8.98	\\
\hline\hline
\multicolumn{6}{|c|}{$\text{M}_{\text{D}} = 3\ \text{TeV}$}\\\hline
Model 1 & $5.59 \times 10^{-5}$	& $3.82 \times 10^{-2}$	& $5.16 \times 10^{-2}$	& $6.58 \times 10^{-2}$	& $1.85 \times 10^{-3}$	\\
\hline
Model 2 & $4.44 \times 10^{-5}$	& $3.04 \times 10^{-2}$ & $4.10 \times 10^{-2}$	& $5.23 \times 10^{-2}$	& $1.47 \times 10^{-3}$	\\
\hline
Model 3 & $3.96 \times 10^{-5}$	& $2.70 \times 10^{-2}$ & $3.65 \times 10^{-2}$	& $4.66 \times 10^{-3}$	& $1.31 \times 10^{-3}$	\\
\hline
Model 4	& $3.74 \times 10^{-5}$	& $2.55 \times 10^{-2}$ & $3.44 \times 10^{-2}$	& $4.40 \times 10^{-3}$	& $1.23 \times 10^{-3}$	\\
\hline
Model 5	& $3.54 \times 10^{-5}$	& $2.42 \times 10^{-2}$ & $3.26 \times 10^{-2}$	& $4.16 \times 10^{-3}$	& $1.17 \times 10^{-3}$	\\
\hline\hline
\multicolumn{6}{|c|}{$\text{M}_{\text{D}} = 5\ \text{TeV}$}\\\hline
Model 1 & $8.50 \times 10^{-9}$ & $5.91 \times 10^{-6}$	& $7.98 \times 10^{-6}$	& $1.02 \times 10^{-5}$	& $2.84 \times 10^{-7}$	\\
\hline
Model 2 & $6.75 \times 10^{-9}$ & $4.69 \times 10^{-6}$ & $6.33 \times 10^{-6}$	& $8.08 \times 10^{-6}$	& $2.25 \times 10^{-7}$	\\
\hline
Model 3 & $7.22 \times 10^{-9}$ & $5.02 \times 10^{-6}$ & $6.78 \times 10^{-6}$	& $8.65 \times 10^{-6}$	& $2.41 \times 10^{-7}$	\\
\hline
Model 4	& $6.82 \times 10^{-9}$ & $4.73 \times 10^{-6}$ & $6.39 \times 10^{-6}$	& $8.16 \times 10^{-6}$	& $2.28 \times 10^{-7}$	\\
\hline
Model 5	& $6.45 \times 10^{-9}$ & $4.48 \times 10^{-6}$ & $6.05 \times 10^{-6}$	& $7.73 \times 10^{-6}$	& $2.16 \times 10^{-7}$	\\
\hline
\end{tabular}
\caption{Cross sections for production of a QBH decaying into an electron and positron with a CoM energy of the initial particles of 8 TeV.\label{EE2}}
\end{center}
\end{table}

\begin{table}[!ht]
\begin{center}
\begin{tabular}{|c|c|c|c|c|c|}\hline
cross section in fb & \multicolumn{5}{c|}{Model for low scale quantum gravity}\\\hline
$\sigma \left(\text{QBH} \rightarrow \text{e}^{+} + \text{e}^{-} \right)$ & 4 dim\cite{Calmet:2008tn}  & ADD n = 5 & ADD n = 6 & ADD n = 7 & RS\\\hline\hline
\multicolumn{6}{|c|}{CoM energy of $\sqrt{\text{s}} = 14\ \text{TeV}$}\\\hline\hline
\multicolumn{6}{|c|}{$\text{M}_{\text{D}} = 1\ \text{TeV}$}\\\hline
Model 1 & 3.25 	& $1.59 \times 10^{3}$ & $2.14 \times 10^{3}$	& $2.72 \times 10^{3}$	& 84.22	\\
\hline
Model 2 & 2.62 	& $1.29 \times 10^{3}$ & $1.73 \times 10^{3}$	& $2.20 \times 10^{3}$	& 68.06	\\
\hline
Model 3 & 2.57	& $1.29 \times 10^{3}$ & $1.73 \times 10^{3}$	& $2.20 \times 10^{3}$	& 67.78\\
\hline
Model 4	& 2.43  & $1.22 \times 10^{3}$ & $1.64 \times 10^{3}$	& $2.09 \times 10^{3}$	& 64.16	\\
\hline
Model 5	& 2.31 	& $1.16 \times 10^{3}$ & $1.56 \times 10^{3}$	& $1.98 \times 10^{3}$	& 60.92	\\
\hline\hline
\multicolumn{6}{|c|}{$\text{M}_{\text{D}} = 3\ \text{TeV}$}\\\hline
Model 1 & $4.43 \times 10^{-3}$	& 2.71	& 3.66	& 4.67	& 0.14	\\
\hline
Model 2 & $3.53 \times 10^{-3}$	& 2.17	& 2.92	& 3.72	& 0.11	\\
\hline
Model 3 & $3.23 \times 10^{-3}$	& 1.99	& 2.68	& 3.41	& $9.97 \times 10^{-2}$	\\
\hline
Model 4	& $3.05 \times 10^{-3}$	& 1.88	& 2.53	& 3.22	& $9.42 \times 10^{-2}$	\\
\hline
Model 5	& $2.89 \times 10^{-3}$	& 1.78	& 2.40	& 3.05	& $8.92 \times 10^{-2}$	\\
\hline\hline
\multicolumn{6}{|c|}{$\text{M}_{\text{D}} = 5\ \text{TeV}$}\\\hline
Model 1 & $3.41 \times 10^{-5}$ & $2.24 \times 10^{-2}$	& $3.03 \times 10^{-2}$	& $3.86 \times 10^{-2}$	& $1.10 \times 10^{-3}$	\\
\hline
Model 2 & $2.71 \times 10^{-5}$ & $1.78 \times 10^{-2}$ & $2.41 \times 10^{-2}$	& $3.07 \times 10^{-2}$	& $8.76 \times 10^{-4}$	\\
\hline
Model 3 & $2.43 \times 10^{-5}$ & $1.60 \times 10^{-2}$ & $2.15 \times 10^{-2}$	& $2.74 \times 10^{-2}$	& $7.83 \times 10^{-4}$	\\
\hline
Model 4	& $2.29 \times 10^{-5}$ & $1.51 \times 10^{-2}$ & $2.03 \times 10^{-2}$	& $2.59 \times 10^{-2}$	& $7.39 \times 10^{-4}$	\\
\hline
Model 5	& $2.17 \times 10^{-5}$ & $1.43 \times 10^{-2}$ & $1.92 \times 10^{-2}$	& $2.45 \times 10^{-2}$	& $7.00 \times 10^{-4}$	\\
\hline
\end{tabular}
\caption{Cross sections for production of a QBH decaying into an electron and positron with a CoM energy of the initial particles of 14 TeV.\label{EE3}
}
\end{center}
\end{table}

\newpage

\begin{table}[!ht]
\begin{center}
\begin{tabular}{|c|c|c|c|c|c|}\hline
cross section in fb & \multicolumn{5}{c|}{Model for low scale quantum gravity}\\\hline
$\sigma \left(\text{QBH} \rightarrow \mu^{+} + \text{e}^{-} \right)$ & 4 dim\cite{Calmet:2008tn}  & ADD n = 5 & ADD n = 6 & ADD n = 7 & RS\\\hline\hline
\multicolumn{6}{|c|}{CoM energy of $\sqrt{\text{s}} = 7\ \text{TeV}$}\\\hline\hline
\multicolumn{6}{|c|}{$\text{M}_{\text{D}} = 1\ \text{TeV}$}\\\hline
Model 2 & 0.20 	& 126.84	& 171.09	& 218.36	& 6.22	\\
\hline
Model 4	& 0.17  & 113.37	& 152.92	& 195.18	& 5.55	\\
\hline
Model 5	& 0.17 	& 107.50	& 145.01	& 185.09	& 5.26	\\
\hline\hline
\multicolumn{6}{|c|}{$\text{M}_{\text{D}} = 3\ \text{TeV}$}\\\hline
Model 2 & $1.01 \times 10^{-5}$	& $6.96 \times 10^{-3}$ & $9.40 \times 10^{-3}$	& $1.20 \times 10^{-2}$	& $3.35 \times 10^{-4}$	\\
\hline
Model 4	& $8.53 \times 10^{-6}$	& $5.89 \times 10^{-3}$ & $7.95 \times 10^{-3}$	& $1.02 \times 10^{-2}$	& $2.84 \times 10^{-4}$	\\
\hline
Model 5	& $8.08 \times 10^{-6}$	& $5.58 \times 10^{-3}$ & $7.53 \times 10^{-3}$	& $9.62 \times 10^{-3}$	& $2.69 \times 10^{-4}$	\\
\hline\hline
\multicolumn{6}{|c|}{$\text{M}_{\text{D}} = 5\ \text{TeV}$}\\\hline
Model 2 & $2.55 \times 10^{-10}$& $1.77 \times 10^{-7}$ & $2.40 \times 10^{-7}$	& $3.06 \times 10^{-7}$	& $8.53 \times 10^{-9}$	\\
\hline
Model 4	& $2.82 \times 10^{-10}$& $1.97 \times 10^{-7}$ & $2.66 \times 10^{-7}$	& $3.39 \times 10^{-7}$	& $9.45 \times 10^{-9}$	\\
\hline
Model 5	& $2.57 \times 10^{-10}$& $1.86 \times 10^{-7}$ & $2.51 \times 10^{-7}$	& $3.21 \times 10^{-7}$	& $8.95 \times 10^{-9}$	\\
\hline
\end{tabular}
\caption{Cross sections for production of a QBH decaying into an electron and anti-muon with a CoM energy of the initial particles of 7 TeV. Due to lepton flavor violation this process is forbidden in model 1 and 3. With the symmetry restrictions in models 2, 4 and 5 the cross sections are identical to the $\text{e}^{+} + \text{e}^{-}$ case.\label{EMU1}
}
\end{center}
\end{table}

\begin{table}[!ht]
\begin{center}
\begin{tabular}{|c|c|c|c|c|c|}\hline
cross section in fb & \multicolumn{5}{c|}{Model for low scale quantum gravity}\\\hline
$\sigma \left(\text{QBH} \rightarrow \mu^{+} + \text{e}^{-} \right)$ & 4 dim\cite{Calmet:2008tn} & ADD n = 5 & ADD n = 6 & ADD n = 7 & RS\\\hline\hline
\multicolumn{6}{|c|}{CoM energy of $\sqrt{\text{s}} = 8\ \text{TeV}$}\\\hline\hline
\multicolumn{6}{|c|}{$\text{M}_{\text{D}} = 1\ \text{TeV}$}\\\hline
Model 2 & 0.34 	& 211.85	& 285.59	& 364.35	& 10.49	\\
\hline
Model 4	& 0.30  & 191.65	& 258.39	& 329.67	&  9.47	\\
\hline
Model 5	& 0.29 	& 181.78	& 245.08	& 312.69	&  8.98	\\
\hline\hline
\multicolumn{6}{|c|}{$\text{M}_{\text{D}} = 3\ \text{TeV}$}\\\hline
Model 2 & $4.44 \times 10^{-5}$	& $3.04 \times 10^{-2}$ & $4.10 \times 10^{-2}$	& $5.23 \times 10^{-2}$	& $1.47 \times 10^{-3}$	\\
\hline
Model 4	& $3.74 \times 10^{-5}$	& $2.55 \times 10^{-2}$ & $3.44 \times 10^{-2}$	& $4.40 \times 10^{-3}$	& $1.23 \times 10^{-3}$	\\
\hline
Model 5	& $3.54 \times 10^{-5}$	& $2.42 \times 10^{-2}$ & $3.26 \times 10^{-2}$	& $4.16 \times 10^{-3}$	& $1.17 \times 10^{-3}$	\\
\hline\hline
\multicolumn{6}{|c|}{$\text{M}_{\text{D}} = 5\ \text{TeV}$}\\\hline
Model 2 & $6.75 \times 10^{-9}$ & $4.69 \times 10^{-6}$ & $6.33 \times 10^{-6}$	& $8.08 \times 10^{-6}$	& $2.25 \times 10^{-7}$	\\
\hline
Model 4	& $6.82 \times 10^{-9}$ & $4.73 \times 10^{-6}$ & $6.39 \times 10^{-6}$	& $8.16 \times 10^{-6}$	& $2.28 \times 10^{-7}$	\\
\hline
Model 5	& $6.45 \times 10^{-9}$ & $4.48 \times 10^{-6}$ & $6.05 \times 10^{-6}$	& $7.73 \times 10^{-6}$	& $2.16 \times 10^{-7}$	\\
\hline
\end{tabular}
\caption{Cross sections for production of a QBH decaying into an electron and anti-muon with a CoM energy of the initial particles of 8 TeV. Due to lepton flavor violation this process is forbidden in model 1 and 3. With the symmetry restrictions in models 2, 4 and 5 the cross sections are identical to the $\text{e}^{+} + \text{e}^{-}$ case.\label{EMU2}}
\end{center}
\end{table}

\begin{table}[!ht]
\begin{center}
\begin{tabular}{|c|c|c|c|c|c|}\hline
cross section in fb & \multicolumn{5}{c|}{Model for low scale quantum gravity}\\\hline
$\sigma \left(\text{QBH} \rightarrow \mu^{+} + \text{e}^{-} \right)$ & 4 dim\cite{Calmet:2008tn}  & ADD n = 5 & ADD n = 6 & ADD n = 7 & RS\\\hline\hline
\multicolumn{6}{|c|}{CoM energy of $\sqrt{\text{s}} = 14\ \text{TeV}$}\\\hline\hline
\multicolumn{6}{|c|}{$\text{M}_{\text{D}} = 1\ \text{TeV}$}\\\hline
Model 2 & 2.62 	& $1.29 \times 10^{3}$ & $1.73 \times 10^{3}$	& $2.20 \times 10^{3}$	& 68.06	\\
\hline
Model 4	& 2.43  & $1.22 \times 10^{3}$ & $1.64 \times 10^{3}$	& $2.09 \times 10^{3}$	& 64.16\\
\hline
Model 5	& 2.31 	& $1.16 \times 10^{3}$ & $1.56 \times 10^{3}$	& $1.98 \times 10^{3}$	& 60.92\\
\hline\hline
\multicolumn{6}{|c|}{$\text{M}_{\text{D}} = 3\ \text{TeV}$}\\\hline
Model 2 & $3.53 \times 10^{-3}$	& 2.17	& 2.92	& 3.72	& 0.11	\\
\hline
Model 4	& $3.05 \times 10^{-3}$	& 1.88	& 2.53	& 3.22	& $9.42 \times 10^{-2}$	\\
\hline
Model 5	& $2.89 \times 10^{-3}$	& 1.78	& 2.40	& 3.05	& $8.92 \times 10^{-2}$	\\
\hline\hline
\multicolumn{6}{|c|}{$\text{M}_{\text{D}} = 5\ \text{TeV}$}\\\hline
Model 2 & $2.71 \times 10^{-5}$ & $1.78 \times 10^{-2}$ & $2.41 \times 10^{-2}$	& $3.07 \times 10^{-2}$	& $8.76 \times 10^{-4}$	\\
\hline
Model 4	& $2.29 \times 10^{-5}$ & $1.51 \times 10^{-2}$ & $2.03 \times 10^{-2}$	& $2.59 \times 10^{-2}$	& $7.39 \times 10^{-4}$	\\
\hline
Model 5	& $2.17 \times 10^{-5}$ & $1.43 \times 10^{-2}$ & $1.92 \times 10^{-2}$	& $2.45 \times 10^{-2}$	& $7.00 \times 10^{-4}$	\\
\hline
\end{tabular}
\caption{Cross sections for production of a QBH decaying into an electron and anti-muon with a CoM energy of the initial particles of 14 TeV. Due to lepton flavor violation this process is forbidden in model 1 and 3. With the symmetry restrictions in models 2, 4 and 5 the cross sections are identical to the $\text{e}^{+} + \text{e}^{-}$ case.\label{EMU3}
}
\end{center}
\end{table}

\newpage

\begin{table}[!ht]
\begin{center}
\begin{tabular}{|c|c|c|c|c|c|}\hline
cross section in fb & \multicolumn{5}{c|}{Model for low scale quantum gravity}\\\hline
$\sigma \left(\text{QBH} \rightarrow \text{u} + \text{e}^{-} \right)$ & 4 dim\cite{Calmet:2008tn}  & ADD n = 5 & ADD n = 6 & ADD n = 7 & RS\\\hline\hline
\multicolumn{6}{|c|}{CoM energy of $\sqrt{\text{s}} = 7\ \text{TeV}$}\\\hline\hline
\multicolumn{6}{|c|}{$\text{M}_{\text{D}} = 1\ \text{TeV}$}\\\hline
Model 4	& $4.17 \times 10^{-2}$ & 28.32	& 38.23	& 48.82	& 1.37	\\
\hline
Model 5	& $3.48 \times 10^{-2}$	& 23.60	& 31.86	& 40.68	& 1.14	\\
\hline\hline
\multicolumn{6}{|c|}{$\text{M}_{\text{D}} = 3\ \text{TeV}$}\\\hline
Model 4	& $3.94 \times 10^{-7}$	& $2.73 \times 10^{-4}$ & $3.69 \times 10^{-4}$	& $4.71 \times 10^{-4}$	& $1.31 \times 10^{-5}$	\\
\hline
Model 5	& $3.28 \times 10^{-7}$	& $2.28 \times 10^{-4}$ & $3.07 \times 10^{-4}$	& $3.92 \times 10^{-4}$	& $1.10 \times 10^{-5}$	\\
\hline\hline
\multicolumn{6}{|c|}{$\text{M}_{\text{D}} = 5\ \text{TeV}$}\\\hline
Model 4	& $8.04 \times 10^{-12}$& $5.60 \times 10^{-9}$ & $7.56 \times 10^{-9}$	& $9.66 \times 10^{-9}$	& $2.69 \times 10^{-10}$	\\\hline
Model 5	& $6.70 \times 10^{-12}$& $4.67 \times 10^{-9}$ & $6.30 \times 10^{-9}$	& $8.05\times 10^{-9}$	& $2.24 \times 10^{-10}$	\\\hline
\end{tabular}
\caption{Cross sections for production of a QBH decaying into an electron and up quark with a CoM energy of the initial particles of 7 TeV. Due to symmetry violations this process is forbidden in models 1, 2 and 3.\label{EU1}
}
\end{center}
\end{table}

\begin{table}[!ht]
\begin{center}
\begin{tabular}{|c|c|c|c|c|c|}\hline
cross section in fb & \multicolumn{5}{c|}{Model for low scale quantum gravity}\\\hline
$\sigma \left(\text{QBH} \rightarrow \text{u} + \text{e}^{-} \right)$ & 4 dim\cite{Calmet:2008tn} & ADD n = 5 & ADD n = 6 & ADD n = 7 & RS\\\hline\hline
\multicolumn{6}{|c|}{CoM energy of $\sqrt{\text{s}} = 8\ \text{TeV}$}\\\hline\hline
\multicolumn{6}{|c|}{$\text{M}_{\text{D}} = 1\ \text{TeV}$}\\\hline
Model 4	& $8.11 \times 10^{-2}$ & 54.07 & 72.96	& 93.16	&  2.63	\\
\hline
Model 5	& $6.76 \times 10^{-2}$ & 45.06 & 60.80	& 77.63	&  2.19	\\
\hline\hline
\multicolumn{6}{|c|}{$\text{M}_{\text{D}} = 3\ \text{TeV}$}\\\hline
Model 4	& $2.23 \times 10^{-6}$	& $1.53 \times 10^{-3}$ & $2.07 \times 10^{-3}$	& $2.64 \times 10^{-3}$	& $7.40 \times 10^{-5}$	\\
\hline
Model 5	& $1.85 \times 10^{-6}$	& $1.28 \times 10^{-3}$ & $1.73 \times 10^{-3}$	& $2.20 \times 10^{-3}$	& $6.17 \times 10^{-5}$	\\
\hline\hline
\multicolumn{6}{|c|}{$\text{M}_{\text{D}} = 5\ \text{TeV}$}\\\hline
Model 4	& $1.91 \times 10^{-10}$& $1.33 \times 10^{-7}$ & $1.79 \times 10^{-7}$ & $2.29 \times 10^{-7}$	& $6.38 \times 10^{-9}$ \\
\hline
Model 5	& $1.59 \times 10^{-10}$& $1.11 \times 10^{-7}$ & $1.49 \times 10^{-7}$	& $1.91 \times 10^{-7}$	& $5.32 \times 10^{-9}$	\\
\hline
\end{tabular}
\caption{Cross sections for production of a QBH decaying into an electron and up quark with a CoM energy of the initial particles of 8 TeV. Due to symmetry violations this process is forbidden in models 1, 2 and 3.\label{EU2}}
\end{center}
\end{table}

\newpage

\begin{table}[!ht]
\begin{center}
\begin{tabular}{|c|c|c|c|c|c|}\hline
cross section in fb & \multicolumn{5}{c|}{Model for low scale quantum gravity}\\\hline
$\sigma \left(\text{QBH} \rightarrow \text{u} + \text{e}^{-} \right)$ & 4 dim\cite{Calmet:2008tn}  & ADD n = 5 & ADD n = 6 & ADD n = 7 & RS\\\hline\hline
\multicolumn{6}{|c|}{CoM energy of $\sqrt{\text{s}} = 14\ \text{TeV}$}\\\hline\hline
\multicolumn{6}{|c|}{$\text{M}_{\text{D}} = 1\ \text{TeV}$}\\\hline
Model 4	& 0.88  & 510.27	& 687.19	& 876.07	& 25.74	\\
\hline
Model 5	& 0.73 	& 425.23	& 572.66	& 730.06	& 21.45 \\
\hline\hline
\multicolumn{6}{|c|}{$\text{M}_{\text{D}} = 3\ \text{TeV}$}\\\hline
Model 4	& $4.30 \times 10^{-4}$	& 0.28	& 0.37	& 0.48	& $1.37 \times 10^{-2}$	\\
\hline
Model 5	& $3.58 \times 10^{-4}$	& 0.23	& 0.31	& 0.40	& $1.14 \times 10^{-2}$	\\
\hline\hline
\multicolumn{6}{|c|}{$\text{M}_{\text{D}} = 5\ \text{TeV}$}\\\hline
Model 4	& $1.44 \times 10^{-6}$ & $9.66 \times 10^{-4}$ & $1.30 \times 10^{-3}$	& $1.66 \times 10^{-3}$	& $4.71 \times 10^{-5}$	\\
\hline
Model 5	& $1.20 \times 10^{-6}$ & $8.05 \times 10^{-4}$ & $1.09 \times 10^{-3}$	& $1.39 \times 10^{-3}$	& $3.93 \times 10^{-5}$	\\
\hline
\end{tabular}
\caption{Cross sections for production of a QBH decaying into an electron and up quark with a CoM energy of the initial particles of 14 TeV. Due to symmetry violations this process is forbidden in models 1, 2 and 3.\label{EU3}
}
\end{center}
\end{table}

\newpage

\begin{table}[!ht]
\begin{center}
\begin{tabular}{|c|c|c|c|c|c|}\hline
cross section in fb & \multicolumn{5}{c|}{Model for low scale quantum gravity}\\\hline
$\sigma \left(\text{QBH} \rightarrow \text{d} + \mu^{+} \right)$ & 4 dim\cite{Calmet:2008tn}  & ADD n = 5 & ADD n = 6 & ADD n = 7 & RS\\\hline\hline
\multicolumn{6}{|c|}{CoM energy of $\sqrt{\text{s}} = 7\ \text{TeV}$}\\\hline\hline
\multicolumn{6}{|c|}{$\text{M}_{\text{D}} = 1\ \text{TeV}$}\\\hline
Model 4	& $5.92 \times 10^{-2}$ & 40.29	& 54.38 & 69.45	& 1.95	\\
\hline
Model 5	& $4.74 \times 10^{-2}$	& 32.23 	& 43.51	& 55.56& 1.56	\\
\hline\hline
\multicolumn{6}{|c|}{$\text{M}_{\text{D}} = 3\ \text{TeV}$}\\\hline
Model 4	& $4.33 \times 10^{-7}$	& $3.00 \times 10^{-4}$ & $4.05 \times 10^{-4}$	& $5.17 \times 10^{-4}$	& $1.44 \times 10^{-5}$	\\
\hline
Model 5	& $3.46 \times 10^{-7}$	& $2.40 \times 10^{-4}$ & $3.24 \times 10^{-4}$	& $4.14 \times 10^{-4}$	& $1.16 \times 10^{-5}$	\\
\hline\hline
\multicolumn{6}{|c|}{$\text{M}_{\text{D}} = 5\ \text{TeV}$}\\\hline
Model 4	& $8.69 \times 10^{-12}$& $6.05 \times 10^{-9}$ & $8.18 \times 10^{-9}$	& $1.04 \times 10^{-8}$	& $2.91 \times 10^{-10}$	\\\hline
Model 5	& $6.95 \times 10^{-12}$& $4.84 \times 10^{-9}$ & $6.54 \times 10^{-9}$	& $8.35\times 10^{-9}$	& $2.33 \times 10^{-10}$	\\\hline
\end{tabular}
\caption{Cross sections for production of a QBH decaying into an anti-muon and down quark with a CoM energy of the initial particles of 7 TeV. Due to symmetry violations this process is forbidden in models 1, 2 and 3.\label{AMD1}}
\end{center}
\end{table}

\begin{table}[!ht]
\begin{center}
\begin{tabular}{|c|c|c|c|c|c|}\hline
cross section in fb & \multicolumn{5}{c|}{Model for low scale quantum gravity}\\\hline
$\sigma \left(\text{QBH} \rightarrow \text{d} + \mu^{+} \right)$ & 4 dim\cite{Calmet:2008tn} & ADD n = 5 & ADD n = 6 & ADD n = 7 & RS\\\hline\hline
\multicolumn{6}{|c|}{CoM energy of $\sqrt{\text{s}} = 8\ \text{TeV}$}\\\hline\hline
\multicolumn{6}{|c|}{$\text{M}_{\text{D}} = 1\ \text{TeV}$}\\\hline
Model 4	& 0.12 			& 77.03 & 103.95& 132.72&  3.74	\\
\hline
Model 5	& $9.20 \times 10^{-2}$ & 61.62 & 83.16	& 106.17&  2.99	\\
\hline\hline
\multicolumn{6}{|c|}{$\text{M}_{\text{D}} = 3\ \text{TeV}$}\\\hline
Model 4	& $2.54 \times 10^{-6}$	& $1.75 \times 10^{-3}$ & $2.36 \times 10^{-3}$	& $3.02 \times 10^{-3}$	& $8.44 \times 10^{-5}$	\\
\hline
Model 5	& $2.03 \times 10^{-6}$	& $1.40 \times 10^{-3}$ & $1.89 \times 10^{-3}$	& $2.41 \times 10^{-3}$	& $6.75 \times 10^{-5}$	\\
\hline\hline
\multicolumn{6}{|c|}{$\text{M}_{\text{D}} = 5\ \text{TeV}$}\\\hline
Model 4	& $2.01 \times 10^{-10}$& $1.40 \times 10^{-7}$ & $1.89 \times 10^{-7}$ & $2.41 \times 10^{-7}$	& $6.73 \times 10^{-9}$ \\
\hline
Model 5	& $1.61 \times 10^{-10}$& $1.12 \times 10^{-7}$ & $1.51 \times 10^{-7}$	& $1.93 \times 10^{-7}$	& $5.38 \times 10^{-9}$	\\
\hline
\end{tabular}
\caption{Cross sections for production of a QBH decaying into an anti-muon and down quark with a CoM energy of the initial particles of 8 TeV. Due to symmetry violations this process is forbidden in models 1, 2 and 3.\label{AMD2}}
\end{center}
\end{table}

\newpage

\begin{table}[!ht]
\begin{center}
\vspace{-2cm}
\begin{tabular}{|c|c|c|c|c|c|}\hline
cross section in fb & \multicolumn{5}{c|}{Model for low scale quantum gravity}\\\hline
$\sigma \left(\text{QBH} \rightarrow \text{d} + \mu^{+} \right)$ & 4 dim\cite{Calmet:2008tn}  & ADD n = 5 & ADD n = 6 & ADD n = 7 & RS\\\hline\hline
\multicolumn{6}{|c|}{CoM energy of $\sqrt{\text{s}} = 14\ \text{TeV}$}\\\hline\hline
\multicolumn{6}{|c|}{$\text{M}_{\text{D}} = 1\ \text{TeV}$}\\\hline
Model 4	& 1.22  		& 708.66		& 954.32		& 1216.58		& 35.78	\\
\hline
Model 5	& 0.98 			& 566.93		& 763.46		& 973.27		& 28.62 \\
\hline\hline
\multicolumn{6}{|c|}{$\text{M}_{\text{D}} = 3\ \text{TeV}$}\\\hline
Model 4	& $5.70 \times 10^{-4}$	& 0.37			& 0.50			& 0.64			& $1.82 \times 10^{-2}$	\\
\hline
Model 5	& $4.56 \times 10^{-4}$	& 0.30			& 0.40			& 0.51			& $1.46 \times 10^{-2}$	\\
\hline\hline
\multicolumn{6}{|c|}{$\text{M}_{\text{D}} = 5\ \text{TeV}$}\\\hline
Model 4	& $1.66 \times 10^{-6}$ & $1.11 \times 10^{-3}$ & $1.50 \times 10^{-3}$	& $1.92 \times 10^{-3}$	& $5.42 \times 10^{-5}$	\\
\hline
Model 5	& $1.33 \times 10^{-6}$ & $8.90 \times 10^{-4}$ & $1.20 \times 10^{-3}$	& $1.53 \times 10^{-3}$	& $4.34 \times 10^{-5}$	\\
\hline
\end{tabular}
\caption{Cross sections for production of a QBH decaying into an anti-muon and down quark with a CoM energy of the initial particles of 14 TeV. Due to symmetry violations this process is forbidden in models 1, 2 and 3.\label{AMD3}}
\end{center}
\end{table}

\newpage
\clearpage


\subsubsection*{Conclusions}

The production cross sections can be found in Tables 10-12 while the cross sections for specific final states with potentially small  Standard Model backgrounds can be found in Tables 13-27. We considered the production of quantum black holes with discrete masses at the Large Hadron Collider within several frameworks for low scale quantum gravity. We study collisions at different center of mass energies (7 TeV, 8 TeV and 14 TeV). We investigate ADD with $n=5,6$ and $7$, Randall and Sundrum's brane world and a 4-dimensional construction proposed in  \cite{Calmet:2008tn}. Several models corresponding to different conservation laws are studied and the Planck scale is taken to be 1 TeV, 3 TeV and 5 TeV. Clearly, the cross sections can be sizable depending on the value of the Planck scale. We find that generally speaking the Large Hadron Collider operating at 7 and 8 TeV is only able to probe a Planck mass in the 1 TeV region. We anticipate that the Standard Model background is likely to be too large to probe a 
higher Planck scale. The 14 TeV Large Hadron Collider could probe the Planck scale up to 3 TeV. Rare events such as final states with a muon and a positron back to back could potentially test a larger Planck scale. Quantifying the reach of the Large Hadron Collider requires a dedicated background study and detector simulation which goes beyond the scope of this article. As expected, the cross sections are smaller than in the continuous mass case \cite{Calmet:2008tn}. However, the possible final states are the same as in the continuous mass case if one assumes that the same quantum numbers, i.e. SU(3)$_c$, U(1) and spin are conserved. It is important to stress that the black holes considered here are rather different since their masses are discrete. They are thus more particle-like than the continuous mass black holes considered in the past. Most final states involve two jets back to back and the background might be large. Final states with two photons or two leptons should be easy to identify at the Large 
Hadron Collider and should be searched for. We emphasize that the limits found in the literature on the production of small black holes do not apply to those considered here. Finally, we point out that it is possible to have more exotic signatures with, e.g., a high transverse momentum muon and a high transverse momentum positron back to back, depending on whether lepton flavor symmetries are conserved or not by quantum gravity. These final states are smoking guns for non-thermal quantum black holes.

\bigskip

{\it Acknowledgments:}
We would like to thank Claire Shepherd-Themistocleous for stimulating discussions.
This work is supported in part by the European Cooperation in Science and Technology (COST) action MP0905 ``Black Holes in a Violent  Universe". The work of X.C. was supported by the STFC grant ST/J000477/1. The work of N.G. is supported by a SEPnet PhD fellowship. One of us, N.G., would like to thank D.~M.~Gingrich for helpful clarifications on his paper \cite{Gingrich:2009hj}.


\bigskip
\newpage








 \begin{thebibliography}{99}


  \bibitem{ArkaniHamed:1998rs} 
  N.~Arkani-Hamed, S.~Dimopoulos and G.~R.~Dvali,
  Phys.\ Lett.\ B {\bf 429}, 263 (1998)
  [hep-ph/9803315].

  \bibitem{Randall:1999ee} 
  L.~Randall and R.~Sundrum,
  Phys.\ Rev.\ Lett.\  {\bf 83}, 3370 (1999)
  [hep-ph/9905221].

  \bibitem{Calmet:2008tn} 
  X.~Calmet, S.~D.~H.~Hsu and D.~Reeb,
  Phys.\ Rev.\ D {\bf 77}, 125015 (2008)
  [arXiv:0803.1836 [hep-th]].
  
\bibitem{Calmet:2010nt} 
  X.~Calmet,
  Mod.\ Phys.\ Lett.\ A {\bf 25}, 1553 (2010)
  [arXiv:1005.1805 [hep-ph]].
  
  
  \bibitem{Eardley:2002re}
  D.~M.~Eardley and S.~B.~Giddings,
  Phys.\ Rev.\  D {\bf 66}, 044011 (2002)
  [arXiv:gr-qc/0201034];


  \bibitem{D'Eath:1992hb}
  P.~D.~D'Eath and P.~N.~Payne,
  Phys.\ Rev.\  D {\bf 46}, 658 (1992);
  Phys.\ Rev.\  D {\bf 46}, 675 (1992);
  Phys.\ Rev.\  D {\bf 46}, 694 (1992).
  
  \bibitem{Hsu:2002bd}
  S.~D.~H.~Hsu,
  Phys.\ Lett.\  B {\bf 555}, 92 (2003)
  [arXiv:hep-ph/0203154].

  \bibitem{Meade:2007sz}
  P.~Meade and L.~Randall,
  JHEP {\bf 0805}, 003 (2008)
  [arXiv:0708.3017 [hep-ph]].
  
  \bibitem{Dimopoulos:2001hw}
  S.~Dimopoulos and G.~L.~Landsberg,
  Phys.\ Rev.\ Lett.\  {\bf 87}, 161602 (2001)
  [arXiv:hep-ph/0106295].
 
\bibitem{Banks:1999gd}
  T.~Banks and W.~Fischler,
  arXiv:hep-th/9906038.
  
\bibitem{Giddings:2001bu}
  S.~B.~Giddings and S.~D.~Thomas,
  Phys.\ Rev.\  D {\bf 65}, 056010 (2002)
  [arXiv:hep-ph/0106219].

\bibitem{Feng:2001ib}
  J.~L.~Feng and A.~D.~Shapere,
  Phys.\ Rev.\ Lett.\  {\bf 88}, 021303 (2002)
  [arXiv:hep-ph/0109106].

\bibitem{Anchordoqui:2003ug}
  L.~A.~Anchordoqui, J.~L.~Feng, H.~Goldberg and A.~D.~Shapere,
  Phys.\ Lett.\  B {\bf 594}, 363 (2004)
  [arXiv:hep-ph/0311365].

\bibitem{Anchordoqui:2001cg}
  L.~A.~Anchordoqui, J.~L.~Feng, H.~Goldberg and A.~D.~Shapere,
  Phys.\ Rev.\  D {\bf 65} (2002) 124027
  [arXiv:hep-ph/0112247].
 
\bibitem{Anchordoqui:2003jr}
  L.~A.~Anchordoqui, J.~L.~Feng, H.~Goldberg and A.~D.~Shapere,
  Phys.\ Rev.\  D {\bf 68}, 104025 (2003)
  [arXiv:hep-ph/0307228].
  
\bibitem{Dai:2007ki} 
  D.~-C.~Dai, G.~Starkman, D.~Stojkovic, C.~Issever, E.~Rizvi and J.~Tseng,
  Phys.\ Rev.\ D {\bf 77}, 076007 (2008)
  [arXiv:0711.3012 [hep-ph]].
  
\bibitem{Calmet:2008dg} 
  X.~Calmet, W.~Gong and S.~D.~H.~Hsu,
  Phys.\ Lett.\ B {\bf 668}, 20 (2008)
  [arXiv:0806.4605 [hep-ph]].
  
  \cite{min,mead,Padmanabhan:au,Calmet:2004mp,Calmet:2005mh,Calmet:2007vb}
   \bibitem{min}
  L.~J.~Garay,
  Int.\ J.\ Mod.\ Phys.\ A {\bf 10},  (1995) 145
  [arXiv:gr-qc/9403008].
  
 
  \bibitem{mead}
   C.~A.~Mead, Phys.\ Rev.\ {\bf 135},  (1964) B849.
   
  \bibitem{Padmanabhan:au}
  T.~Padmanabhan,
  Class.\ Quant.\ Grav.\  {\bf 4},  (1987) L107.
  
  \bibitem{Calmet:2004mp} 
  X.~Calmet, M.~Graesser and S.~D.~H.~Hsu,
  Phys.\ Rev.\ Lett.\  \textbf{93}, (2004) 211101 
  [arXiv:hep-th/0405033].
 
   \bibitem{Calmet:2005mh}
  X.~Calmet, M.~Graesser and S.~D.~H.~Hsu,
  Int.\ J.\ Mod.\ Phys.\ D \textbf{14} (2005) 2195
  [arXiv:hep-th/0505144].
  
  \bibitem{Calmet:2007vb} 
  X.~Calmet,
  Eur.\ Phys.\ J.\ C {\bf 54}, 501 (2008)
  [hep-th/0701073].
 
 
\bibitem{Calmet:2012cn} 
  X.~Calmet, D.~Fragkakis and N.~Gausmann,
  ``Non thermal small black holes,''
  arXiv:1201.4463 [hep-ph], in ``Black Holes: Evolution, Theory and Thermodynamics," Eds A.~ J.~Bauer and  D.~G.~Eiffel, Nova Science Publishers,  New York, (2012), 165-170;
  Eur.\ Phys.\ J.\ C {\bf 71}, 1781 (2011)
  [arXiv:1105.1779 [hep-ph]].

  \bibitem{Dvali:2011nh} 
  G.~Dvali, C.~Gomez and S.~Mukhanov,
  arXiv:1106.5894 [hep-ph].
  


  


  \bibitem{Yoshino:2002tx}
  H.~Yoshino and Y.~Nambu,
  Phys.\ Rev.\  D {\bf 67}, 024009 (2003)
  [arXiv:gr-qc/0209003].
 \bibitem{Yoshino2}
  H.~Yoshino and V.~S.~Rychkov,
  Phys.\ Rev.\  D {\bf 71}, 104028 (2005)
  [Erratum-ibid.\  D {\bf 77}, 089905 (2008)]
  [arXiv:hep-th/0503171].
  
  \bibitem{pdg}
  K.~Nakamura et al. (Particle Data Group), 
  J. Phys. G {\bf 37}, 075021 (2010).

\bibitem{Gingrich:2009hj} 
  D.~M.~Gingrich,
  J.\ Phys.\ G G {\bf 37}, 105108 (2010)
  [arXiv:0912.0826 [hep-ph]].
 
\end{thebibliography}
\end{document}